\documentclass[letterpaper, 10 pt, conference]{IEEEtran}  

\usepackage{amsmath} 
\usepackage{amssymb}  

\usepackage{epsfig}

\usepackage{eucal} \usepackage{graphicx} \usepackage{subfig} \usepackage{url,picture}

\usepackage{colordvi,verbatim}

\newcommand{\ls}[1] {\dimen0=\fontdimen6\the\font \lineskip=#1\dimen0
\advance\lineskip.5\fontdimen5\the\font \advance\lineskip-\dimen0
\lineskiplimit=.9\lineskip \baselineskip=\lineskip \advance\baselineskip\dimen0
\normallineskip\lineskip \normallineskiplimit\lineskiplimit
\normalbaselineskip\baselineskip \ignorespaces } 
\newcommand{\Lo} {\Lambda_{out}}\newcommand{\Li} {\Lambda_{in}}\newcommand{\Ld} {\Lambda_{d}}
\newcommand{\ik}[1]{I_{_{#1}}}
\newcommand{\no}{\nonumber}

\newcommand{\SH}[1]{{\mathcal S}(#1)}
\newcommand{\CH}[1]{{\mathcal C}(#1)}
\IEEEoverridecommandlockouts

\newtheorem{proposition}{Proposition}[section]

\newtheorem{remark}{Remark}[section] 

\title{\LARGE \bf Optimal Control in Two-Hop Relay Routing}

\author{\centering Eitan Altman$^\star$\thanks{$^\star$INRIA B.P.93, 2004 Route des Lucioles, 06902 Sophia-Antipolis, Cedex, FRANCE, @-mail: {\tt\small eitan.altman@sophia.inria.fr}.}, Tamer Ba\c{s}ar$^{\dagger}$\thanks{$^{\dagger}$University of Illinois, 1308 West Main Street, Urbana, IL 61801-2307, USA, @-mail: {\tt\small tbasar@control.csl.uiuc.edu}.}, and Francesco De Pellegrini$^{\diamond}$\thanks{$^{\diamond}$CREATE-NET, via Alla Cascata 56 c, 38100 Trento, ITALY, @-mail: {\tt\small fdepellegrini@create-net.org}.}}

\begin{document} 

\date{} \maketitle


\thispagestyle{empty}

\begin{abstract}
We study  the optimal control of propagation of packets
in delay tolerant mobile ad-hoc networks. We consider a two-hop
forwarding policy under which the expected number of nodes carrying
copies of the packets obeys a linear dynamics. We exploit this property to
formulate the  problem in the framework of linear quadratic optimal
control which allows us to obtain closed-form expressions for the optimal
control and to study numerically the tradeoffs  by varying
various parameters that define the cost.
\end{abstract}


\section{Introduction}


In DTN (Delay Tolerant Network) mobile ad-hoc networks, connectivity is not needed
any more and packets can arrive at their destination thanks to the mobility of
some subset of nodes that carry copies of a packet.  A naive approach
to forward a packet to the destination is by epidemic routing in which
any mobile that has the packet keeps on relaying it to any other
mobile that arrives within its transmission range. 
This leads to minimization of
the delivery delay  at a cost, however, of inefficient use of network resources
(in terms of memory used in the relaying mobiles and in terms of the
energy used for flooding the network).
The need for  more efficient use of network resources
motivated the use of the more economic two-hop routing protocols
in which the source transmits copies of its message to all mobiles
it encounters, but where the latter relay the message only if they
come in contact with the destination. Furthermore, timers
have been proposed to be associated with messages when stored
at relay mobiles, so that after some threshold (possibly random)
the message is discarded. The performance of the two-hop
forwarding protocol along with the effect of the timers have
been evaluated in \cite{ANA} which 
brings up the possibility of optimization of the choice
of the average timer duration.

The optimization of the two-hop relay routing as well as of extensions
that we propose here are the central objectives of our present
work. Instead of using fixed parameters, however
(e.g. timers which have fixed average durations
whose values can be determined as in \cite{ANA}),  we propose a dynamic
optimization approach based on  optimal control. Thus the
parameters of the two-hop relay protocol associated with a
message of a given age are allowed to change with time.




DTNs attracted recently the attention of the
networking community~\cite{Levine_infocom2006,Diot_infocom2006,fall_implementingDTN}.
Among DTNs, a relevant case is that of mobile ad-hoc networks, including
 systems where human mobility is used to diffuse information through portable
devices \cite{Diot_infocom2006}.

Lack of persistent connectivity makes routing the central issue
in DTNs \cite{fall_routDTNs,Zegura_Mobihoc06}. The problem is to deliver messages to destinations
with high probability despite the fact that encounter patterns of mobile devices are unknown. Message diffusion
algorithms trade off message delay for energy consumption, e.g., number of copies per delivered message and/or 
transmission range. As mentioned before, the naive solution is epidemic routing \cite{VB}.

The two-hop routing protocol considered here was introduced by Gr{\"o}ssglauser
and Tse in \cite{tse_mobility02}; the main goal there was to characterize the capacity of mobile ad-hoc networks and
the two-hop protocol was meant to overcome severe limitations of static networks capacity obtained in \cite{gupta_capacity}.
Two-hop routing, in particular, provides a convenient compromise of energy versus delay compared to epidemic routing:
the standard reference work for the analysis of the two-hop relaying protocol is \cite{GNK}.
Fluid approximations and infection spreading models similar to those we use here are described extensively
in \cite{ZNKT}. Interestingly, we show (Remark  \ref{rem:meanfield}) that the fluid mean field approximation turns  out to 
be an exact description of the dynamics of the expectation of the system's state due to the linearity of the 
dynamics under two-hop routing.

Algorithms to control forwarding in DTNs have been proposed in the recent literature, e.g,
\cite{MM}, \cite{FBS}. In \cite{MM}, the authors describe an epidemic
forwarding protocol based on the {\em susceptible-infected-removed} (SIR) model
\cite{ZNKT}. They show that it is possible to increase the
message delivery probability  by tuning the parameters of the underlying SIR
model. In \cite{FBS} a detailed general framework is proposed in order to
capture the relative performances of different self-limiting strategies.
Finally, under a fluid model approximation, the work in \cite{ABD} provides a general
framework for the optimal control of the broad class of monotone relay strategies,
i.e., policies where the number of copies do not decrease over time. It is proved there that
optimal forwarding policies are of threshold type.

In this paper, we consider non-monotone relay strategies for 
two-hop routing, and apply optimal control theory to capture general
trade-offs on message delay, energy and storage. After presenting the general 
model in the next section, we formulate
the control problem in Section \ref{contT} and  then provide the solution in
Section \ref{sol-finite}. In Section \ref{sol-infinite} we present
and solve a similar control problem defined on an infinite horizon.
An extension of the initial control problem is studied in Section \ref{DT-control}.  
Section \ref{sec:num} presents a numerical exploration, and 
Section \ref{sec:conclusion} concludes the paper.


\section{The Model}

Consider $K$ classes of mobiles, where class $k$ has
$N_k$ mobiles. Let $N$ be a $K$ dimensional column vector whose $k$-th entry is
$N_k$.  The time between contacts of any two nodes of respective classes $i$
and $j$ is assumed to be exponentially distributed  with some
parameter $\lambda_{ij}$.\footnote{Note that our multidimensional description of the system allows 
one in particular to extend exponentially distributed inter contact times to the much larger class 
of phase type distributions.} The validity of such a model in the special case of a
single class has been discussed in \cite{GNK}, and its accuracy has been shown
for a number of mobility models (Random Walker, Random Direction, Random
Waypoint). 
Let $\Lambda$  be the $K\times K$ matrix whose $ij$-th entry is $\lambda_{ij}$.
We assume that the message that is transmitted is relevant for some duration
$ \tau$. We do not make any assumption on as to whether the source or the
mobiles know whether the messages have reached
the destination within that period or not. 

Let the source be of class $s$ and the destination of class $d$.  Let $
\xi (t)$ be
a $K$-dimensional column vector whose $j$-th entry denotes the size of
population of class $j$ that has the packet.
Let $X(t)= E[ \xi (t)]$.  We assume that each component of $X$ evolves according to
\begin{equation}\label{dyn0}
\frac{ d X_i (t) }{ dt } =\Lambda_{si}(N_i- X_{i}(t)) - \overline M_i(t) X_i(t) \,, i=1,\ldots, K,
\end{equation}
where $\overline M_i(t) \geq 0$. The above dynamics is a generalization of the well-known dynamics of two-hops 
routing (see for example \cite{ABD}), containing the additive  linear control term on $X_i$; this new term represents the effect of 
timeouts by which message copies are discarded at intermediate relays of class $i$. We should point out that, as it will be 
clear later, these dynamics of different components of $X$ will in fact be coupled
by the introduction of a  control term which will be picked optimally, as one minimizing a particular cost function 
that involves the entire vector $X$.
In what follows, we now employ an 
equivalent expression for (\ref{dyn0}) which simplifies the analytic derivation, i.e., 
\begin{equation}\label{dyn1}
\frac{ d X_i (t) }{ dt } =\Lambda_{si}N - \Lambda_{id}X_i(t) - M_i(t) X_i(t) \,, i=1,\ldots, K
\end{equation}
where $M_i(t)=(\Lambda_{si}-\Lambda_{id} +  \overline M_i(t))$. 
For future use, we introduce the compact 
notation 
\[
\frac{ d X (t) }{ dt } =\Lambda_{in}  N- \Lambda_{out} X(t) - M X(t) \,,
\]
where we define $\Lambda_{in}=\mbox{diag}(\Lambda_{s1},\ldots,\Lambda_{sK})$, 
$\Lambda_{out}=$ $\mbox{diag}(\Lambda_{1d},\ldots,\Lambda_{Kd})$, and $M=$$\mbox{diag}(M_1,\ldots,M_K)$. 

Assume that the source has a message at time $0$ and let  $T_d$ denote the amount by which the message is delayed.
Let ${\cal F}_t $ be the $\sigma$-algebra generated by $ \xi_s, s \leq t $.
Denote the conditional successful delivery probability by $ \Psi
(t):= P( T_d < t | {\cal F}_t )$. Given ${\cal F}_t$, the number of
arrivals at the destination during time interval $[0,t]$ is a
Poisson random variable with parameter \( \sum_{i=1}^K \Lambda_{id}
\int_{s=0}^t \xi_i(s) ds . \) Therefore
\begin{equation}
\Psi  (t) = 1 -  \exp \Big( -
\sum_{i=1}^K \Lambda_{id} \int_{s=0}^t \xi_i(s) ds  \Big)
\label{delaySS}
\end{equation}
(An alternative more detailed derivation is given in the Appendix.)
Since $\exp(-x)$ is concave, we have by Jensen's inequality
\begin{equation}
E [ \Psi (t) ] \geq D(t) := 1 - \exp \Big( -
\sum_{i=1}^K \Lambda_{id} \int_{s=0}^t X_i(s) ds  \Big)
\label{delayS}
\end{equation}
\begin{remark}\label{rem:meanfield}
We note that related models using linear differential equations
have frequently been used in DTNs
to approximate the dynamics of the properly
scaled number of mobiles that have a copy of the file; this
is the limit of the mean field dynamics. This type of approximation
has been shown to be tight in various related models
as the total number of nodes $N$
tends to infinity  see e.g. \cite{bcfh}.
As we just saw, it turns out that in the special case of
two hop routing, the dynamics of the mean field limit 
coincides with that of the expectation. On the other hand,
in the mean field limit, the inequality in (\ref{delayS}) becomes
an equality.
\end{remark}


\section{Controlling the timers}
\label{contT}


We may control the timers by allowing $M$ to vary in time.
Let $\widehat u(t)= - M(t) \xi (t)$ and  $u(t)= E [ \widehat u(t) ] $.
We then have:
\begin{equation}
\frac{ d X (t) }{ dt } =  \Lambda_{in}  N- \Lambda_{out} X(t)  + u(t)
\label{dyn2a}
\end{equation}
\begin{equation}
\label{const1}
\noindent\mbox{where}\hskip5mm
0 \leq X(t) \leq N , \quad  u(t) \leq 0\,,
\end{equation}
with the vector inequalities interpreted componentwise.

\noindent{\bf The cost.} The following performance measures could appear as part
of the overall objective:
\begin{enumerate}
\item
{\bf Delay cost:}
We wish to maximize the lower bound given in (\ref{delayS})
on the success probability, i.e. on the probability
that the delay $T_d$ does not exceed the
threshold $\tau$ beyond which the message is considered irrelevant.
Hence we wish to maximize
$ \widetilde X ( \tau ) $, or some monotone function of it,
where $\widetilde X(t) := \sum_{j=1}^K \Lambda_{jd}
\int_{s=0}^t  X_j(s) ds  $.
\item
{\bf Indirect delay cost:}
We may wish
to include a penalty for large values of $u$ (which correspond
to timers that expire at a high rate) as 
higher values of $u$ may result in longer delays.
\item
{\bf Memory cost:}
We may wish to minimize or  bound the number of copies
in the system in order to avoid saturation of the memory available
at the mobiles.
This can be done by directly including a cost on each
of the components $ \widehat X_j(\tau) $ where
$\widehat X_j(t) := \int_0^t X_j(s) ds$ (or having an
instantaneous penalty on $X_j(t)$).
\item
{\bf Energy cost for the network:} The energy cost for class $j$
mobiles is given by
$\;\;
 {\cal E}_0 \Lambda_{jd} \widehat X_j (t)
$.
\end{enumerate}
In view of the above, we introduce the following cost function
corresponding to an initial state $x$ and a policy $u$:
\[
V (t;x,u)=
- c_1 \widetilde X(t)^2
+ c_2 \sum_{j=1}^K \int_0^t  (u_j (s) - \overline u)^2  ds
\]
\[
+ c_3 \sum_{j=1}^K \widehat X_j(t)^2
+ c_4 \sum_{j=1}^K \left( \Lambda_{jd} \widehat X_j(t) \right)^2
\]
where $c_i$'s are all positive.
Let $ R:= - c_1 \Lambda_d \Lambda_d^T + c_3 I + c_4 \Lambda^2_{out} $,
where $\Lambda_d$
is the column vector whose $i$-th entry is given by $\Lambda_{id}$.
Then we can write
\begin{equation}
\label{lqc}
V(t;x,u)=
\sum_{i=1} c_2 \int_0^t  (u_i (s) - \overline u)^2  ds +
\widehat X (t)^T R \widehat X (t)
\end{equation}
We assume henceforth that $R$ is positive semi-definite, 
and we can also take $c_2=1$ without any loss of generality.

The objective is then to minimize $V ( \tau;0,u)$.

By state augmentation, we have a standard optimal control problem with quadratic cost (\ref{lqc}) and 
linear state dynamics:
\[
\frac{ d X (t) }{ dt }  =  - \Lambda_{out} X(t) + u(t) +
\Lambda_{in} N , \ \
\frac{ d \widehat X (t) }{ dt } = X(t)
\]
The state and control are restricted according to (\ref{const1}).

One could use the theory of constrained linear quadratic control, such
as \cite{BMDP}. Or, one could choose to track a value that is sufficiently
far from the boundary so that a controller without the
constraints (as those in (\ref{const1}))  will be satisfied in practice
with a high probability. 


\section{Solution to the Optimal Control Problem}
\label{sol-finite}


\noindent
Let $Z( t) :=  ( X_1(t) , ... , X_K (t) , \widehat X_1 (t) , ... , \widehat X_K (t) )^T $.
Then, the  composite state dynamics can be written as:
$$
{dZ \over dt} = A Z + Bu + \tilde{c} ,
\mbox{ where  }
A = \left( \begin{array}{cc}
-\Lambda_{out} &  0 \\ & \\ I &  0 \end{array} \right) ,
$$
$$
Z = \left(
\begin{array}{c}
X \\ \widehat{X}
\end{array}
\right) , \quad
B=\left( \begin{array}{c} I \\  0 \end{array} \right) ,  \quad
\tilde{c} = \left( \begin{array}{c}
\Lambda_{in} N \\ 0  \end{array} \right)\,,
$$
and the cost function (expressed in terms of $Z$, and with terminal time $\tau$) becomes:
$$
J(\tau; Z, u) = Z^T(\tau) Q_f Z(\tau) + \int^{\tau}_0 (u-\bar{u})^T
(u-\bar{u}) dt
$$
where  with $R\geq0$  defined as earlier, {\small $Q_f = \left( \begin{array}{cc} 0 & 0 \\ & \\ 0 & R \end{array} \right)$}.
Letting $w:= u-\bar{u}$, we can rewrite the above as
$$
{dZ \over dt} = A Z + Bw + c
\mbox{ where }
c:= \left( \begin{array}{c} \Lambda_{in} N + \bar{u} \\ 0 \end{array} \right)
$$
$$J(\tau; Z, w) = Z^T(\tau) Q_f Z(\tau) + \int^{\tau}_0 w^T w dt
$$
Hence what we have is an affine-quadratic optimal control problem \cite{BO}. For 
each fixed finite $\tau$, this  problem admits a
{\em unique strongly time-consistent} optimal solution:
\[
w(t) = -B^T (PZ + k)
\]
where
$P(t) \geq 0, k(t)$ are unique continuously differentiable solutions of
$$
\dot{P} + PA + A^TP - PBB^TP = 0,\quad P(\tau) = Q_f
$$
$$\dot{k} + A^T k + Pc - PBB^Tk = 0,\quad  k(\tau) = 0
$$
$$
\mbox{and, } \quad \min J = Z(0)^T P(0)Z(0) + 2 k^T(0)Z(0) + 2 m(0)
$$
where $m$ is the unique solution of
$$
\dot{m} + k^Tc - {1\over 2} k^TBB^Tk = 0,\quad m(\tau) = 0
$$
With this solution at hand, one can of course readily
obtain the expression for optimal $u = w+\bar{u}$, and
solve for the trajectory of $Z$, and hence of $X$,  but
only numerically.
We should also note that there is no guarantee in the solution above that 
$\bar{u} < u(t) < 0 \; \forall t$, and $0 < X(t) < N$; this can only be verified numerically.


\section{Infinite-Horizon Control for Evolving Files}
\label{sol-infinite}


We consider in this section the transmission of evolving files, that is
 files whose contents  evolve and change from time to time.
The source wishes to send the file to the destination and also send updates
from time to time. The source need not know when the file changes.
Updates of the file may thus be  transmitted at times that are independent
from the instants when the file changes.
Some examples are:
\begin{itemize}
\item
A source has a file containing update information such as
weather forecast or news headlines. 
\item
A source makes backups of some directories and store them at other nodes 
in order to improve  reliability.
\end{itemize}

The information received becomes less relevant as
time passes.  As in the  original model, a relay node
activates a time-to-live (TTL) timer when it receives a packet and deletes
the packet when the timer expires
as there is little interest in relaying old information. %

We are now interested in guaranteeing that there will always be packets in the system
(as recent as possible) so that updated versions could be received
at the destinations. The time horizon is now infinite so we have
to restrict to those components of the cost defined in Section
\ref{contT} which do not depend on the end of the horizon.

We wish to have $X$ large (close to $N$) in order to have a small
delivery delay (in view of (\ref{delayS})). On the other hand we
shall assign cost for low $u_i$ to avoid old information to be
relayed to the destination. 

We let here $Z:=X-N$, and obtain the corresponding state dynamics
$$
{dZ \over dt} = -\Lambda_{out} Z + w + (\Lambda_{in} -
\Lambda_{out})N + \bar{u}
$$

Cost function (with $Q= {\rm diag} (q_1, \ldots , q_K), q_i > 0$):
$$
J(\infty ; Z, w) = \int^\infty_0Z^T(t) Q Z(t) dt +
\int^\infty_0 w^T w dt
$$
$$
\equiv
\sum^K_{i=1} \int^\infty_0 (q_i z_i^2 + w_i^2)dt
$$
This is a completely decoupled problem, whose solution
involves solutions of $K$ scalar optimal control problems.
The $i$-th problem is:
\[
\min J_i = \int^\infty_0 (q_iz_i^2 + w_i^2) dt,
\]
\[
{dz_i\over dt} = -\lambda_i z_i + w_i + (\mu_i-\lambda_i)N_i + \bar{u}_i
\]
where $\lambda_i $ is the $ii$-th element of $\Lambda_{out}$,
and $\mu_i$ is the $ii$-th element of $\Lambda_{in}$. 

Unique stabilizing solution is (we drop the indices,
and hence this solution is for the generic case, with everything being scalar):
$$
w=-(pz+k),\quad -2p\lambda - p^2 + q = 0,
$$
$$-
\lambda k + p(\mu N - \lambda N + \bar{u}) - pk = 0$$
$$
\Rightarrow \;\, p= -\lambda + \sqrt{\lambda^2 + q},
\quad k = p(\mu N - \lambda N + \bar{u})  / \sqrt{\lambda^2 + q}
$$
Under this control policy, the $i$-th state dynamics is:
$$
\dot{x}_i = - \lambda_i x_i + u_i + \mu_iN_i,
$$
$$
u_i = \bar{u}_i -p_i \big(x_i - N_i +
{\bar{u}_i + (\mu_i - \lambda_i)N_i \over \lambda_i + p_i}\big)
$$
which is stable because $\lambda_i + p_i = \sqrt{\lambda^2_i + q_i} > 0$.
The steady-state value of $x_i$ can be obtained by setting the
derivative equal to zero in the expression above, leading to:
$$
x^\infty_i = {1\over \sqrt{\lambda^2_i + q_i} }\bigg( \mu_i +
p_i\big(
1-{\mu_i-\lambda_i)\over \sqrt{\lambda^2_i + q_i} }\big)\bigg) N_i +
{\lambda_i \over
\sqrt{\lambda^2_i + q_i} } \bar{u}_i
$$
and the steady-state value of control is
$ u^\infty_i = \lambda_i x_i^\infty - \mu_i N_i\,.
$
What remains to be shown is that there exists a choice of
$q_i$ under which the bounds on $x_i$ and $u_i$ are met.

As an arbitrary special case, we picked $\lambda_i = 3,
q_i = 16$ (which led to $p_i=2$), and found that as long as
$
\mu_i < 3 - {\bar{u}_i \over N_i}
$
both bounds are met. That is, $0< x^\infty_i< N_i$ and
$\bar{u}_i < u^\infty_i < 0$.

\begin{remark}
if we want the optimal control
to be linear (and not affine) in $x_i$,
we start with
$$
u_i = \bar{u}_i -p_i \Big(x_i - N_i + {\bar{u}_i + (\mu_i - \lambda_i)N_i \over \lambda_i + p_i}\Big)
$$
and add to the right-hand-side the following term which is identically zero on the optimum
trajectory at steady state (where $\alpha_i$ is a scalar parameter):
$
-\alpha_i ( x_i - x^\infty_i ) .
$
Now pick $\alpha_i$ such that all terms other than $x_i$ are zero
(and there is a unique such $\alpha_i$), leaving us with
$
u_i = -(p_i + \alpha_i) x_i
$.
\end{remark}

\begin{remark}  The above analysis can be
extended to the case where there is a running
cost on $\widehat{X}$, but then depending on the
structure of this cost, decoupling may no
longer be possible. Still, LQR theory would
be applicable here. With coupling, it may
not be possible to show that the bounds are
satisfied (only through numerical
computation and simulation).
\end{remark}

\section{Discrete-Time Control}\label{DT-control}


We shall now  assume that the controlled parameters are updated
periodically rather than continuously. Since the time scales
involved in DTN networks are between minutes to hours, this
is expected not to have much effect on the performance and
yet it would decrease computing (and thus energy) resources.
We first consider here the discrete-time version of the finite-horizon problem
of Section~\ref{sol-finite}.


\subsection{Finite-horizon control in discrete time}
\label{DT-finite}


With uniform sampling at every $\Delta$ units of time, the discrete-time version of the state equation
for $Z$ introduced in Section~\ref{sol-finite} is
$$Z_{\ell+1} = F_\ell Z_\ell + \tilde{B}_\ell u_\ell + \tilde{n}_\ell$$
where $Z_\ell$ is $Z(t_\ell)$, with $t_\ell$ being the $\ell$-th sampling time, and $t_{\ell+1} - t_\ell = \Delta$;
$u_\ell= u(t_\ell)$, with control held constant over the subinterval $[t_\ell, t_{\ell+1})$; and $F_\ell= \Phi(t_{\ell+1}, t_\ell)$,
where $\Phi$ satisfies (and is the unique solution of) the matrix differential equation
$${d\Phi(t, \tau) \over dt} = A \Phi(t,\tau)\,,\;\; \Phi(\tau, \tau) = I$$
that is, it is the state transition matrix associated with $A$. Furthermore,
$$\tilde{B}_\ell := \int^{t_{\ell+1}}_{t_\ell} \Phi(t_{\ell+1}, \tau)d\tau\, B\,,\quad
\tilde{n}_\ell := \int^{t_{\ell+1}}_{t_\ell} \Phi(t_{\ell+1}, \tau)d\tau\, \tilde{c}
$$
We note that if matrix $\Lambda_{out}$ was a constant matrix (not time dependent),
then $A$ would be a  constant matrix, and $\Phi(t_{\ell+1}, t_\ell)$ would depend only on
$\Delta$, and likewise $\tilde{B}_\ell $ and
$\tilde{n}_\ell$, which would be constants.
$\Phi(t_{\ell+1}, t_\ell)$ 
can be computed to be
$$\Phi(t_{\ell+1}, t_\ell)  = \left( \begin{array}{cc}
Y&0 \\ \Lambda_{out}^{-1}
[I-Y]&I\end{array} \right)\,,\; Y:= \exp{(-\Lambda_{out} \Delta)}\,,$$
and
$$\tilde{B}_\ell = \left( \begin{array}{cc}
\Lambda_{out}^{-1} [I-Y]&0\\r
\Lambda_{out}^{-1}\big[
\Delta I -\Lambda_{out}^{-1} [I-Y]\big] & I\Delta\end{array} \right) B$$
$$\tilde{n}_\ell = \left( \begin{array}{cc}
\Lambda_{out}^{-1} [I-Y]&0\\
\Lambda_{out}^{-1}\big[
\Delta I -\Lambda_{out}^{-1} [I-Y]\big] & I\Delta\end{array} \right)
\tilde{c}$$

The expression for $F_\ell$ can be further simplified (approximately) if $\Delta > 0$ is very small. To first
order in $\Delta$, $Y= I - \Lambda_{out} \Delta$, and hence
$$
F_\ell = F = \left( \begin{array}{cc} I&0 \\
I\Delta & I\end{array} \right)\,,\quad \tilde{B}_\ell= \tilde{B} = \left( \begin{array}{cc} I\Delta \\
2 \Lambda_{out}^{-1}  \Delta \end{array} \right),
$$
$$
\tilde{n}_\ell = \tilde{n} = \tilde{B} \Lambda_{in} N$$
\smallskip
Now the cost function, again as the counterpart of the one in Section~\ref{sol-finite}, is
\begin{small}
$$
J(L; Z, u) = Z^T_{L+1} Q_f Z_{L+1} + \sum^{L}_{\ell=1} (u_\ell-\bar{u})^T(u_\ell-\bar{u})
+  \sum^{L}_{\ell=1}Z^T_{\ell} QZ_{\ell}
$$
\end{small}
where $Q_f$ is as before, and we have included an additional cost on intermediate values of $Z$, 
with nonnegative definite weighting matrix $Q$ (which can also be taken to be zero). In relation 
to the continuous-time problem, here $L$ is picked such that $t_{L+1} = \tau$.

As before, introducing the new control variable, $w_\ell = u_\ell - \bar{u}$, the state equation becomes
$$Z_{\ell+1} = F_\ell Z_\ell + \tilde{B}_\ell w_\ell + {n_\ell}\,,\quad
n_\ell:= \tilde{n}_\ell  + \tilde{B}_\ell \bar{u}\,,$$ and the cost
function is equivalently
$$
J(L; Z, u) = Z^T_{L+1} Q_f Z_{L+1} + \sum^{L}_{\ell=1} w_\ell^T w_\ell + \sum^{L}_{\ell=1}Z^T_{\ell} QZ_{\ell}
$$
This is a standard linear-quadratic optimal control problem, which admits a 
{\em unique strongly time-consistent} optimal solution \cite{BO}, given by
$$w_\ell = -P_\ell S_{\ell+1}F_\ell Z_\ell - P_\ell (s_{\ell+1} + S_{\ell+1} n_\ell)\,,\;\; \ell=1, 2, \ldots, L\,,$$
where $S_\ell$ and $s_\ell$ are generated by the recursive matrix equations
$$P_\ell=[I+\tilde{B}_\ell^TS_{\ell+1}\tilde{B}_\ell ]^{-1} \tilde{B}_\ell^T$$
$$S_\ell = Q + F_\ell^T S_{\ell+1} [I - \tilde{B}_\ell P_\ell S_{\ell+1}] F_\ell\,,\;\; S_{L+1}=Q_f$$
$$s_\ell= F_\ell^T [I - \tilde{B}_\ell P_\ell S_{\ell+1}] ^T [s_{\ell+1}
+ S_{\ell+1} n_\ell]\,,\;\; s_{L+1} = 0$$ Again, with this solution at
hand, one can  readily  obtain the expression for the optimal $u =
w+\bar{u}$, and  generate  the trajectory of $Z$, and hence of $X$.


\subsection{Infinite-horizon control in discrete time for evolving files}
\label{DT-infinite}


We now obtain the discrete-time counterpart of the result of Section~\ref{sol-infinite},
for the direct  discrete-time version of the model of that section. Following the arguments
of the previous subsection, the state dynamics in discrete time are (assuming that $\Lambda_{out}$
and $\Lambda_{in} $ have time-invariant, constant entries, and $\delta$ is as defined earlier):
$$ Z_{\ell + 1} = G Z_\ell + \bar{B} w_\ell + \tilde{N}\,,\quad \ell = 1, 2, \ldots $$
$$
\mbox{where }
G := \exp{(-\Lambda_{out} \Delta)} = \mbox{diag}\left(e^{-\lambda_i\Delta}\right)$$
$$\bar{B} := \Lambda_{out}^{-1}\left[ I - \exp{(-\Lambda_{out} \Delta)} \right]$$
$$\tilde{N} := \bar{B} [ (\Lambda_{in} - \Lambda_{out} ) N + \bar{u} ]$$
The cost function is (again $Q= {\rm diag} (q_1, \ldots , q_K), q_i > 0$):
$$
J(\infty ; Z, w) = \sum^\infty_{\ell = 1} \left( Z^T_\ell Q Z_\ell +
 w^T_\ell w_\ell \right) \equiv
\sum^K_{i=1} \sum^\infty_{\ell = 1}  (q_i z_{i\ell}^2 + w_{i\ell}^2)
$$

This is again a completely decoupled problem, whose solution
involves solutions of $K$ scalar optimal control problems.
The $i$-th problem is:
$$
\min J_i = \sum^\infty_{\ell = 1}  (q_i z_{i\ell}^2 + w_{i\ell}^2),
\qquad z_{i(\ell) + 1}= g_i z_{i \ell}+ b_iw_{i\ell} + n_i
$$
where
$$ g_i := e^{-\lambda_i \Delta}\,,\;\; b_i := {1\over \lambda_i} (1-g_i)\,,\;\; n_i := b_i[(\mu_i-\lambda_i)N_i + \bar{u}_i]$$
and as before $\lambda_i $ is the $ii$-th element of $\Lambda_{out}$,
and $\mu_i$ is the $ii$-th element of $\Lambda_{in}$.

The unique stabilizing solution is (we drop the indices,
and hence this solution is for the generic case, with everything being scalar):
$$
w=-PSgz - P(s+Sn),\quad P= {b\over 1+Sb^2},
$$
$$
 S=q+ g^2S[1-bPS],\quad  s=g[1-bPS][s-Sn],$$
$$
\Rightarrow \qquad S= {1\over 2b^2}\left[{b^2q + g^2 - 1 + \sqrt{(b^2q + g^2 - 1)^2 + 4qb^2}}\right],
$$
$$
\quad s+Sn = {1+b^2S\over 1+b^2S -g}Sn
$$
where it can be shown that $1+b^2S -g\not= 0$, and hence the expressions for $S$ and $s$ are well
defined. The control policy simplifies to
$$ w = - {bSg\over 1+Sb^2} z - {bS\over 1-g+b^2S}n$$
which is again for the generic case (for the $i$-th control all quantities above will have the index $i$).
Under this control policy, the $i$-th state dynamics is:
$$
z_{i(\ell + 1)} =  {g_i\over 1+ S_ib_i^2} z_{i\ell} + {1-g_i \over 1-g_i + S_ib^2_i}n_i$$
which is stable because $0 < g_i / (1+S_ib^2_i) < 1$, with the reason being that $0 < g_i < 1$ and
$1+S_ib^2_i > 1$. The steady-state value of $z_i$ is:
$$z^\infty_i = {(1-g_i) (1+S_ib^2_i) \over (1-g_i + S_ib^2_i)^2} n_i$$
$$
\mbox{where  }
 n_i = {1\over \lambda_i} (1-e^{\lambda_i\Delta}) [(\mu_i - \lambda_i)N + {\bar{u}}_i]$$
and the actual state is of course $x^\infty_i  = z^\infty_i  + N$. The steady-state open-loop value of
the optimal control is
$$u^\infty_{i} = {\bar{u}}_i  - {b_iS_ig_i\over 1+S_ib_i^2} z^\infty_i - {b_iS_i\over 1-g_i+b_i^2S_i}n_i$$

If $\Delta$ is very small, asymptotically
$S_i\Delta \to -\lambda_i + \sqrt{q_i + \lambda_i^2}$
and the optimal control policy becomes
$$u^{opt}_i = - \left[-\lambda_i + \sqrt{q_i + \lambda_i^2}\right]x_i + {\lambda_i \over \sqrt{q_i +
\lambda_i^2}}{\bar{u}}_i
$$
$$
- \left[(1- {\lambda_i \over \sqrt{q_i + \lambda_i^2}})(\mu_i-\lambda_i) +
\lambda_i - \sqrt{q_i + \lambda_i^2}\right]N
$$
Again in the asymptotic case (for small $\Delta$), the
steady-state value of the state ($i$-th component) is
$$x^\infty_{i}  = N + {\lambda_i (\mu_i-\lambda_i)N\over \sqrt{q_i + \lambda_i^2}}\Delta + {\lambda_i
\over \sqrt{q_i + \lambda_i^2}}{\bar{u}}_i \Delta$$ Note that as $\Delta \to 0$,  $x^\infty_{i}  \to N$.


\section{Numerical Results}
\label{sec:num}


Here we present numerical results for the dynamic control of timers; in particular, we report on
the dynamics of the system in some reference cases. The numerical integration procedure was performed as follows
\begin{itemize}
\item[1)] the solution $P$ of the differential matrix Riccati equation is calculated over the interval $[0,t_f]$;
\item[2)] the dynamics of $k$ have been obtained integrating backwards the related differential equation;
\item[3)] finally, an interpolated version of vector $k$ and matrix $P$ are input to the forward integration
of the ODE describing the state trajectory $X$;
\item[4)] the control law $u$ and the corresponding message delivery probability are derived accordingly.
\end{itemize}

For the cases described in the following we employed the Matlab ODE suite. Mobile nodes intermeeting 
intensities are dimensioned using the Random Waypoint (RWP) synthetic mobility model for which
intermeeting intensities can be calculated \cite{GNK}. The reference intermeeting intensity,
for the following experiments, is $\lambda_0=2.7875\cdot 10^{-4}$; this value is experienced by RWP mobile nodes
moving on a square playground of side $L=1000$ m, speed $v=4$ m/s and radio range $20$ m.
\begin{figure}[t]
\centering
\includegraphics [width=6.5cm]{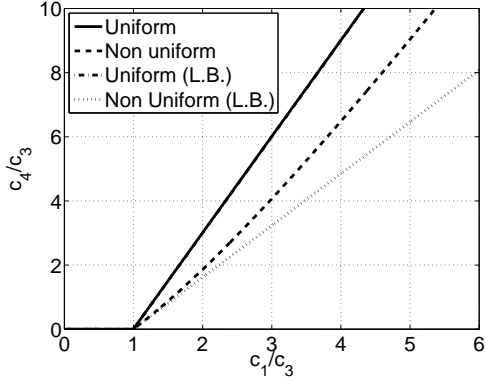}
\caption{The minimum value $c_4/c_3$ as a function of $c_1/c_3$ such that $R>0$; the solid line refers to the uniform case
  $\Ld=(1,1,1)/\sqrt 3$, $\Lo=\mbox{diag}\{(1,1,1)\}$, the dot-dashed line to the non-uniform case
$\Ld=(1,3,5)/\sqrt{35}$, $\Lo=\mbox{diag}\{(1,2,4)\}$.}\label{fig:param}
\end{figure}

\subsection*{Choice of parameters of the linear quadratic optimization}

We first study  the parameters of the optimization problem and related constraints. In particular,
 $c_1, c_3$ and $c_4$ enable to tune the optimization
based on specific weighs that we can assign to the delay, energy and the memory; we notice that 
$c_1,c_3$ and $c_4$ appear in the definition of $R$, whereas $c_2$
appears in the Hamiltonian matrix $H=\left( \begin{array}{cc} A & -BB^T/c_2 \\ 0 & -A^T \end{array} \right)$. 
Note also that we had set $c_2 = 1$ at the outset, as part of normalization of the cost. 

In particular, $c_1,c_3$ and $c_4$ are such that 
\[
R=-c_1\Ld \Ld^t + c_3 I + c_4 \Lo^2>0
\]
In view of the expression for $R$, for a given value of $c_1/c_3$, a minimum value for $c_4/c_3$ exists
such that $R>0$. Here, we provide a simple sufficient condition.
\begin{proposition}\label{prop:suffcond}
 Let \[
\alpha=\frac{\sum_{i=1}^{K} \Lambda_{id}^2}{\sum_{i=1}^{K} \Lambda_{id}^4}\,,
\]
if $c_4 > \alpha (c_1  - c_3)$, then  $R>0$.
\end{proposition}
The proof of the above statement is reported in the Appendix. 
We observe that for $c_1<c_3$ and $c_4>0$ the condition is satisfied. We would also like to give some 
insight in the relative dimensioning of $c_1/c_3$ and $c_4/c_3$ for the outer region $c_1>c_3$. We derived numerically 
the minimum values of
$c_4/c_3$ above for which $R>0$ in case $K=3$. As depicted in Fig.~\ref{fig:param} we reported both on the case of
uniform entries for $\Ld$ and $\Lo$ and the case when they are different; we referred to the normalized case
when $||\Ld||=1$ for the sake of the clearness. There, we can distinguish the inner region for $c_1/c_3<1$
where Prop.~\ref{prop:suffcond} holds, and the outer region where $c_4/c_3$ must increase in order to ensure 
$R>0$; for the tested values the graph shows a quasi linear shape and the lower bound proves 
tight in the case of uniform entries (L.B. lines in Fig.~\ref{fig:param}).

This means that at the increase of $c_1/c_3$, i.e., for larger relative weight given to the delivery probability in the 
cost function, $c_4/c_3$ has to increase: the relative weight given to the energy cost has to increase in order 
to maintain the problem positive definite: this corresponds to the intuition that it is not possible to overweight 
the delivery probability of a message against the energy expenditure.

\subsection*{The effect of the energy constraint ($c_4$)}

\begin{figure*}[t]
\centering
  \subfloat[]{\includegraphics [width=6.5cm]{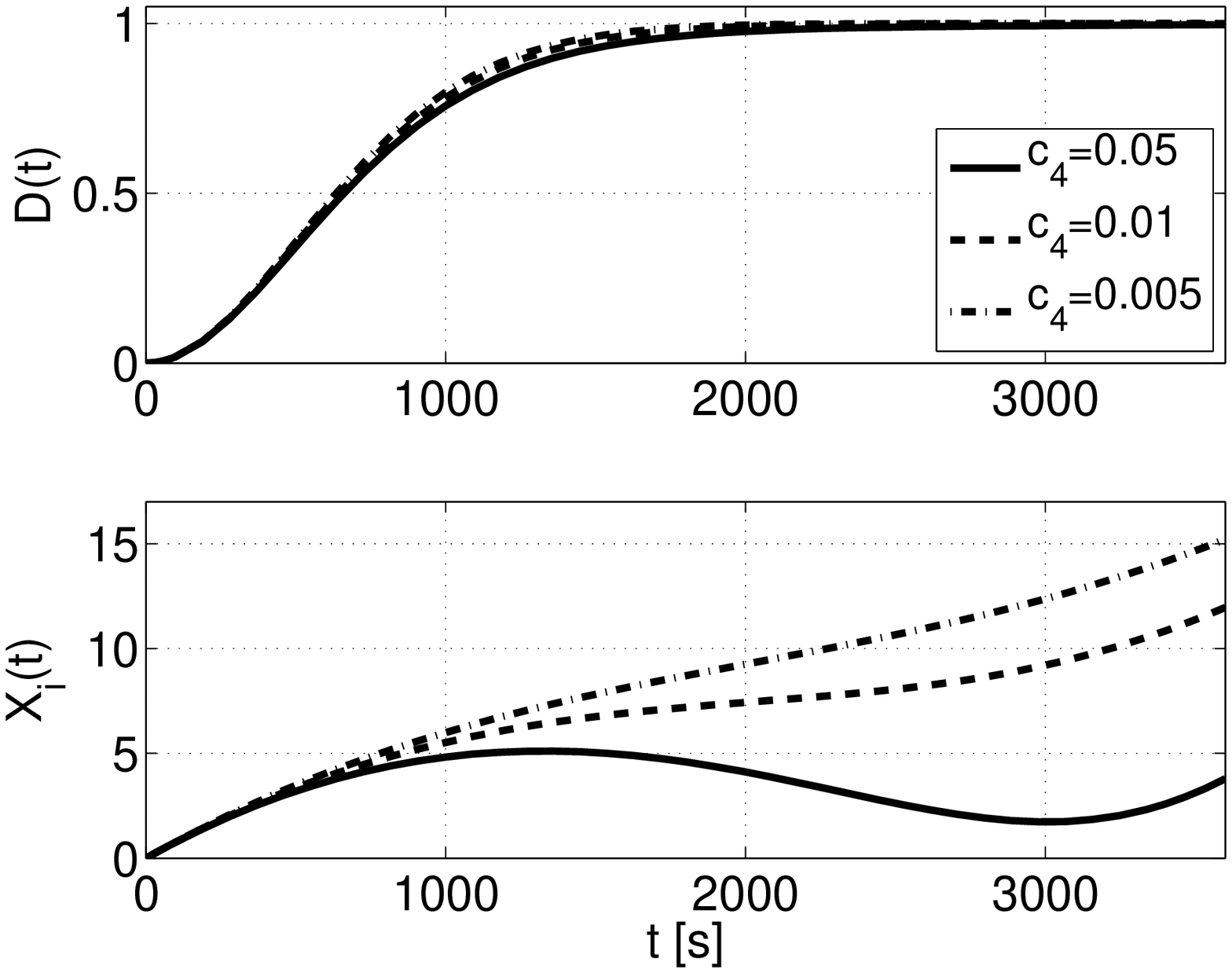}}
  \subfloat[]{\includegraphics [width=6.5cm]{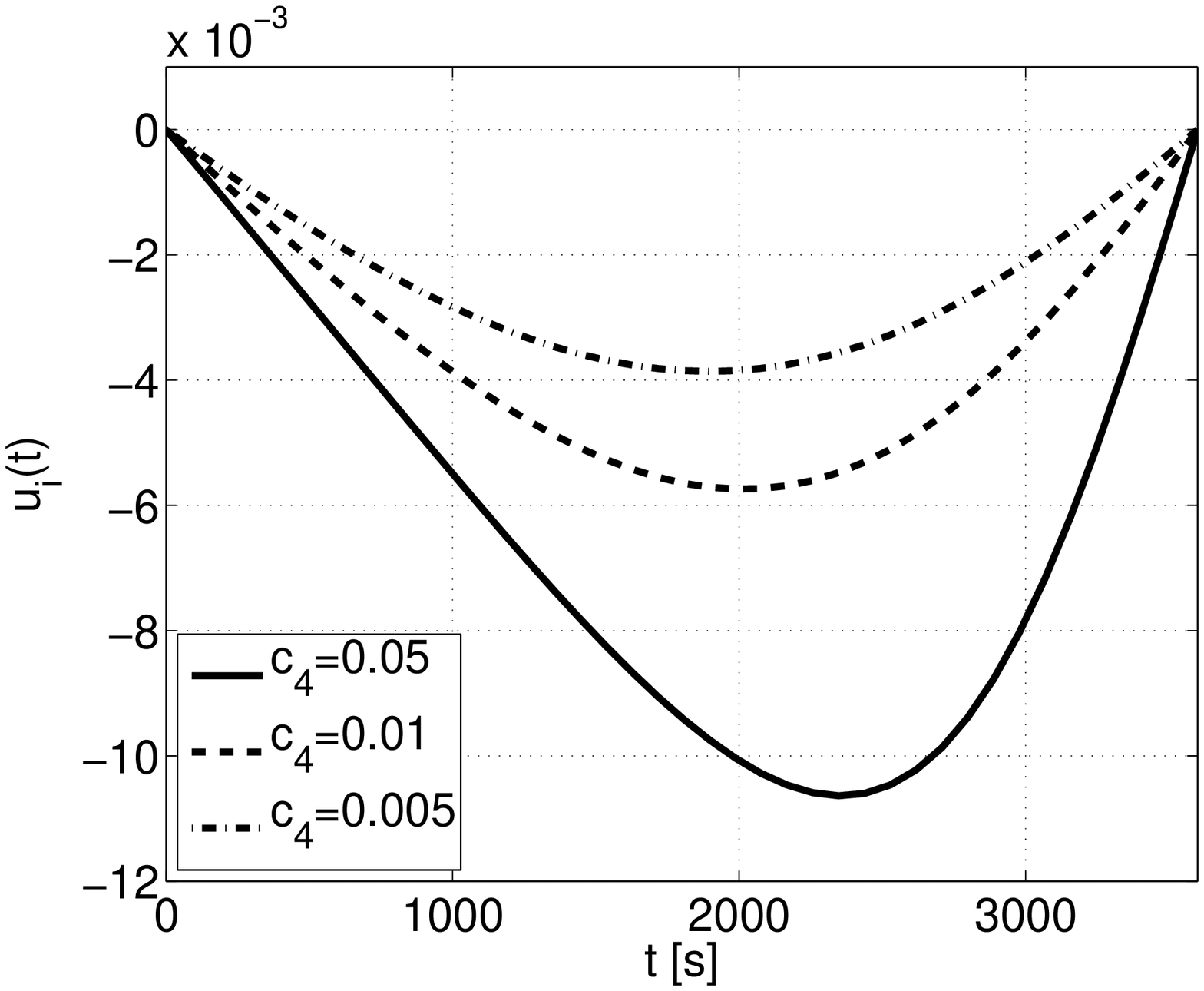}}
\caption{Dynamical control of the timers, uniform case, $N=151$, $\tau=3600$ s; (a) delay CDF (upper) and dynamics
of infected nodes per class $X_i(t)$ (lower); (b) control $u_i$. Effect of $c_4$.%
}\label{fig:uniform_c4}
\end{figure*}

\begin{figure*}[t]
\centering
  \subfloat[]{\includegraphics [width=6.5cm]{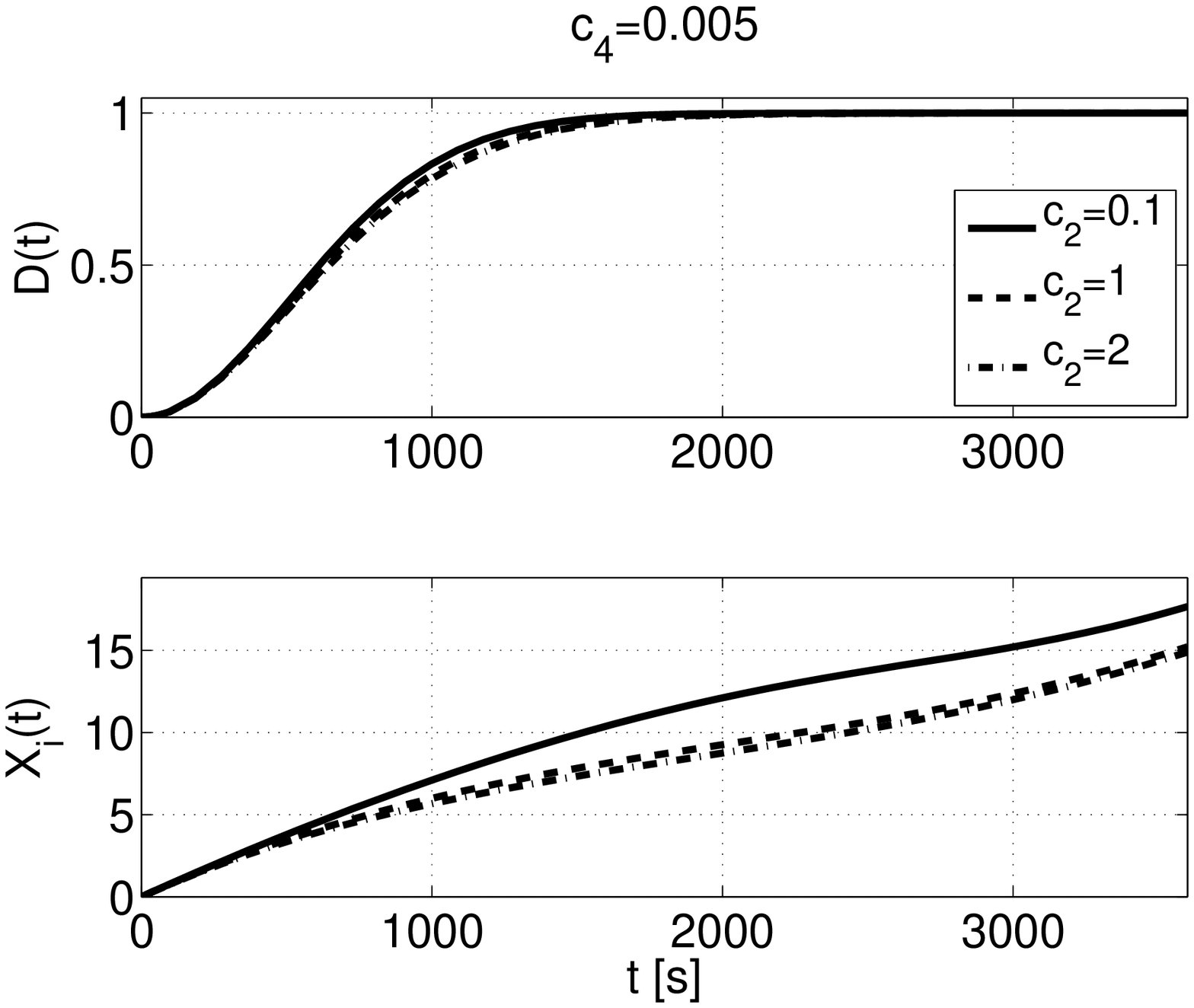}}
  \subfloat[]{\includegraphics [width=6.5cm]{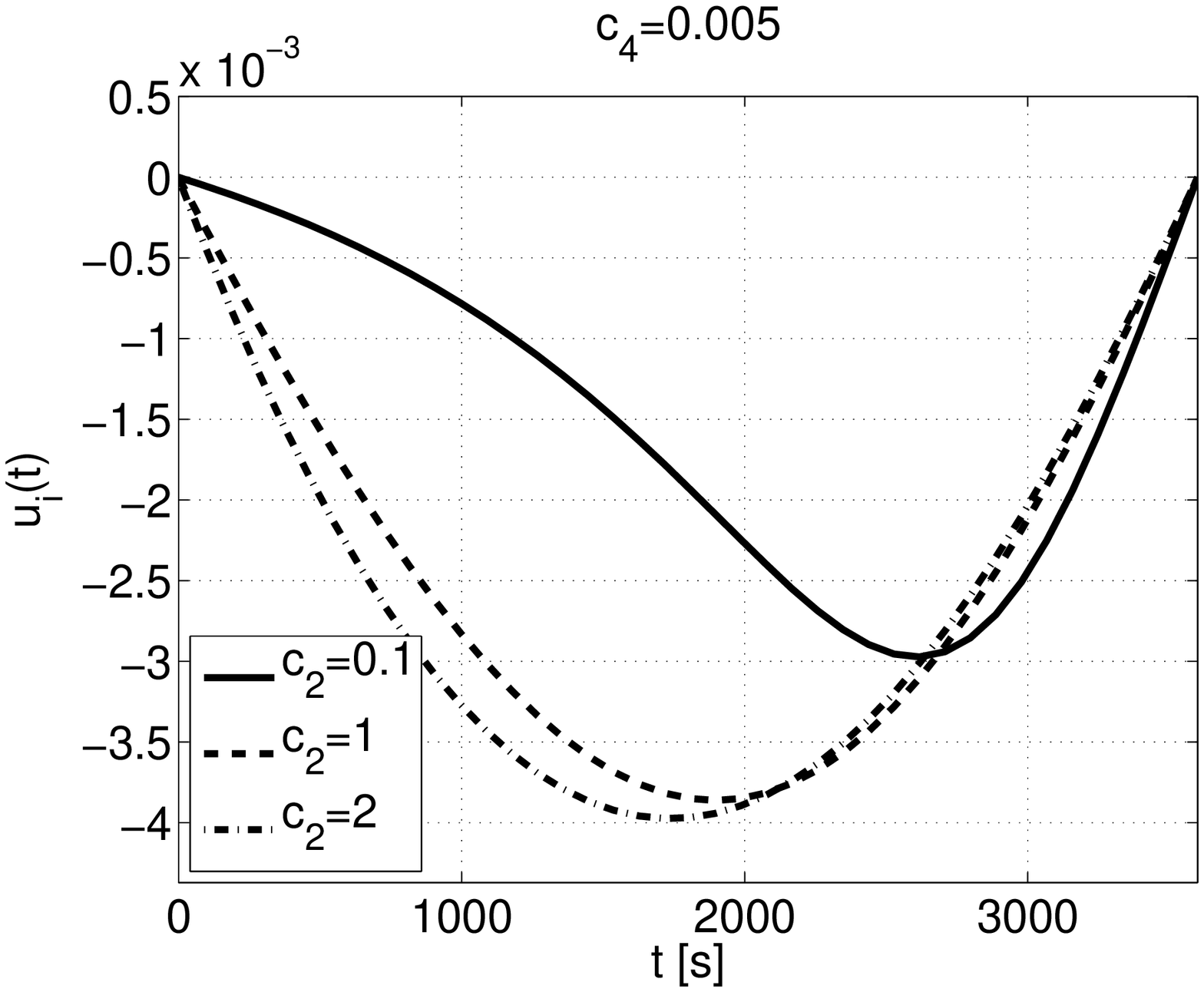}}
\caption{Dynamical control of the timers, uniform case, $N=151$, $\tau=3600$ s; (a) delay CDF (upper) and dynamics of
infected nodes per class $X_i(t)$ (lower); (b) control $u_i$. Effect of $c_2$.%
}\label{fig:uniform_c2_small}
\end{figure*}
First, we describe  the impact of the constraint on energy $c_4$; in particular, we considered terminal 
time $\tau=3600$ s. $N_1-1=N_2=N_3=50$. Also, we let $\overline u=0$.

The coefficients of the optimization problem are $c_1=1/\lambda_0^2$ and $c_3=1$. 
We considered uniform intermeeting intensities, $\lambda_{si}=\lambda_{jd}=\lambda_{0}$, for 
 $i=1,2,3$. 

We compared the performance of the system for three values of $c_4$: $0.05$, $0.01$ and $0.005$. As expected, see
Fig.~\ref{fig:uniform_c4}, when $c_4=0.005$, i.e., the constraint on the energy expenditure for
message transmission is smaller, the effect of timers is milder. For this setting, the message delay CDF reaches the unitary
value much before the terminal time. The controlled dynamics of the number of infected nodes are monotonic as in the case of
the plain, uncontrolled, two-hop routing.

Conversely, for larger values, i.e., $c_4=0.01$, the message delay CDF reaches the unitary value
slightly later than in the previous case (see Fig.~\ref{fig:uniform_c4}a), since the effect of timers becomes
more relevant  (see Fig.~\ref{fig:uniform_c4}b). Furthermore, the change of convexity of the infected nodes dynamics
(see Fig.~\ref{fig:uniform_c4}a) is marked. For $c_4=0.05$, the effect of the larger weight given to the
energy results in a non-monotonic dynamics in the number of copies in the system. We notice that, at the opposite
ends, the stronger constraint on the energy leads to a much smaller number of copies in the system, $10$ per class
in case of $c_4=0.005$, and $4$ in case of $c_4=0.05$: nevertheless in the latter still $D(\tau)\simeq 1$. This
first result already indicates that via proper parameter weighting we can achieve consistent savings of network 
resources.

\subsection*{The case of inhomogeneous classes}

So far we investigated the properties of the system in the case when the classes of mobiles
are homogeneous. Here, 
$\lambda_{1d}=\lambda_{2d}/2=3\lambda_{3d}/2$, whereas the
intermeeting intensity with the source $\lambda_{si}=\lambda_0$, i=1,2,3. In order to maintain consistency
with the previous cases, we normalized the intermeeting intensities $||\Lo||=||\Li||$.
Also, in order to meet the constraints, for this choice of the parameters, a suitable setting is $c_4=0.0025$, $c_3=1/2$, 
$c_1=1/2\lambda_0^2$, and $\overline u_i=-0.004$, $i=1,2,3$.

\begin{figure*}[t]
\centering
  \subfloat[]{\includegraphics [width=6.5cm]{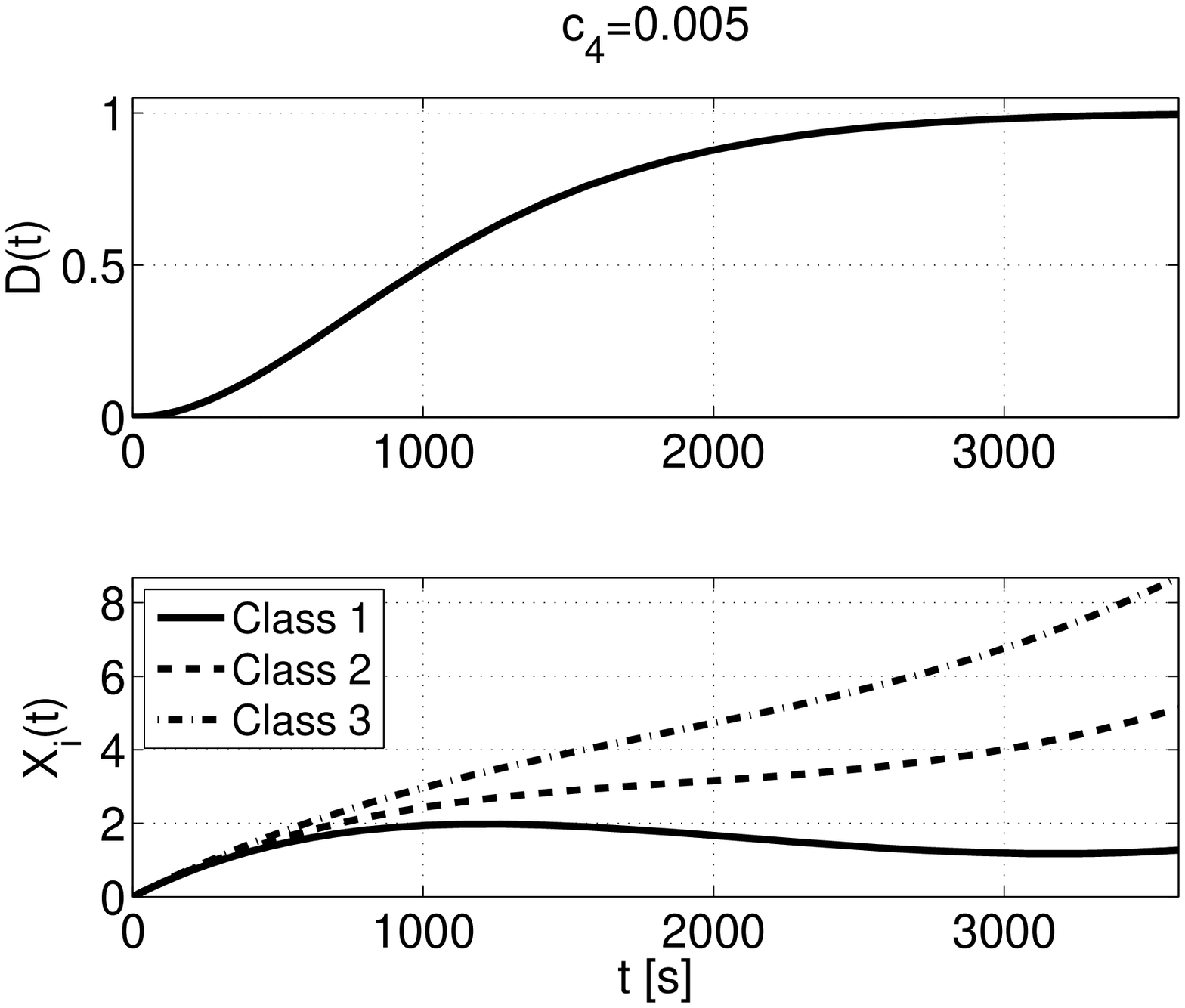}}
  \subfloat[]{\includegraphics [width=6.5cm]{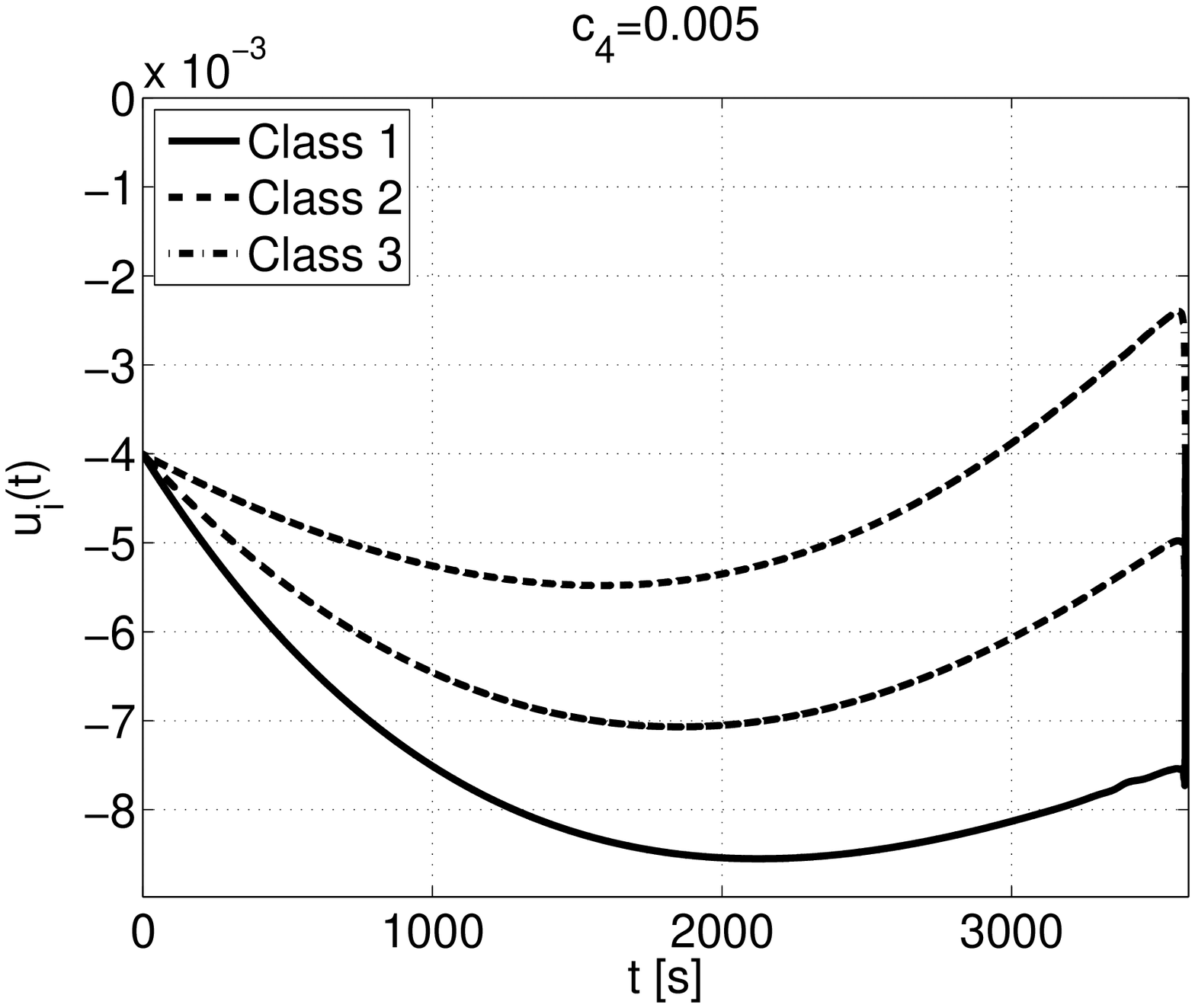}%
             \put(-100,100){\includegraphics [width=2.5cm]{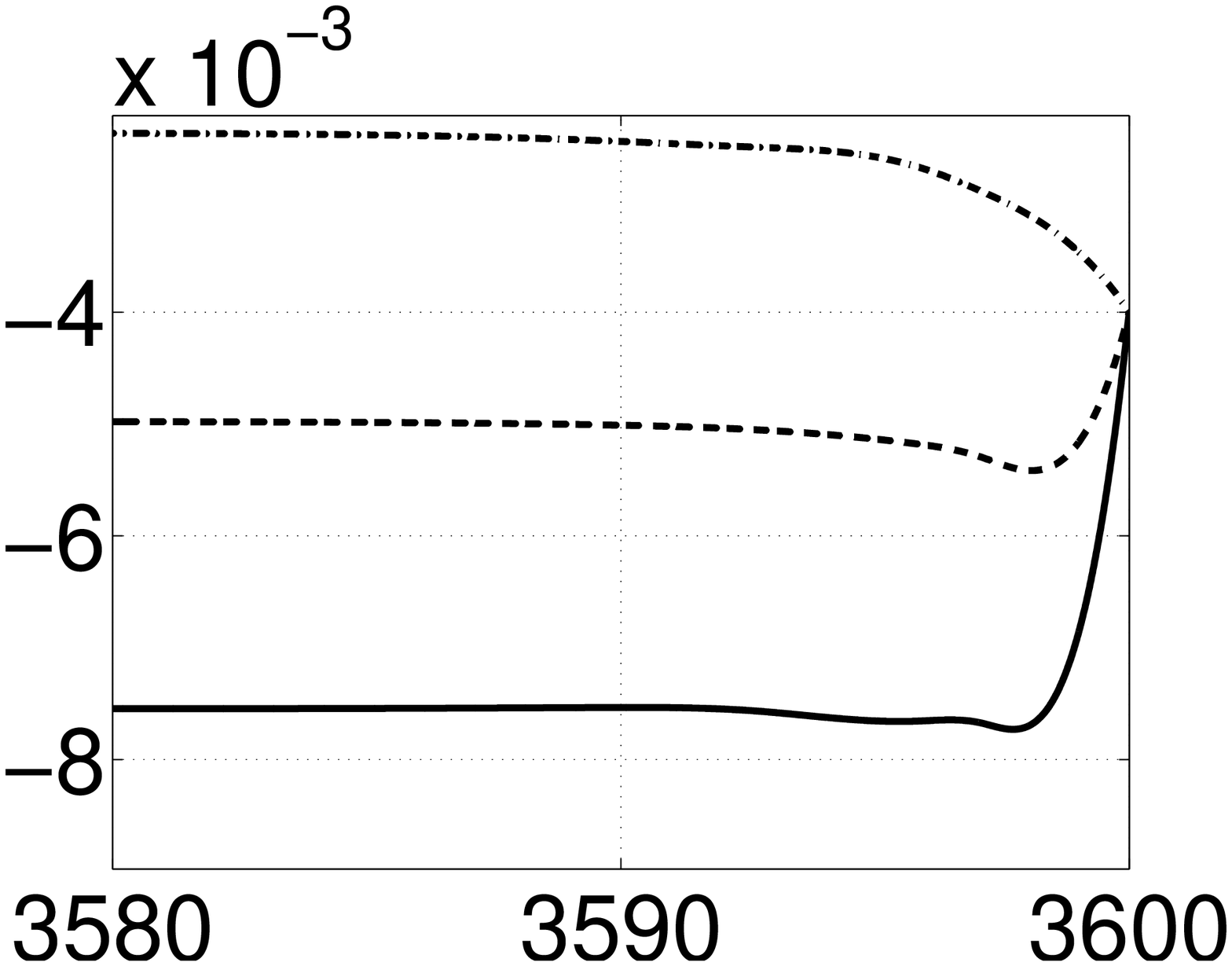}
                           \put(-2,-2){\vector(1,-1){9mm}}}
}
\caption{Dynamical control of the timers, non-uniform case, $N=151$, $\tau=3600$ s; (a) delay CDF (upper) and dynamics
of infected nodes per class $X_i(t)$ (lower); (b) control $u_i$. Parameters are $c_4=0.0025$, $c_3=1/2$, $c_2=1$
and $c_1=1/2\lambda_0^2$. Intermeeting intensities with destination $\lambda_{1d}=1/2 \lambda_{2d}=3/2 \lambda_{3d}$,
$||\Lo||=||\Li||=\lambda_0$.%
}\label{fig:nonuniform_O}
\end{figure*}

In this numerical evaluation, see Fig.~\ref{fig:nonuniform_O}, 
the control 
discriminates classes with higher chances to deliver the message (classes $2$ and $3$) and the first class; 
hence, higher timer rates are assigned to the first class to limit the number of
copies forwarded to nodes with lower intermeeting rates to the destination. 

\begin{figure*}[t]
\centering
  \subfloat[]{\includegraphics [width=6.5cm]{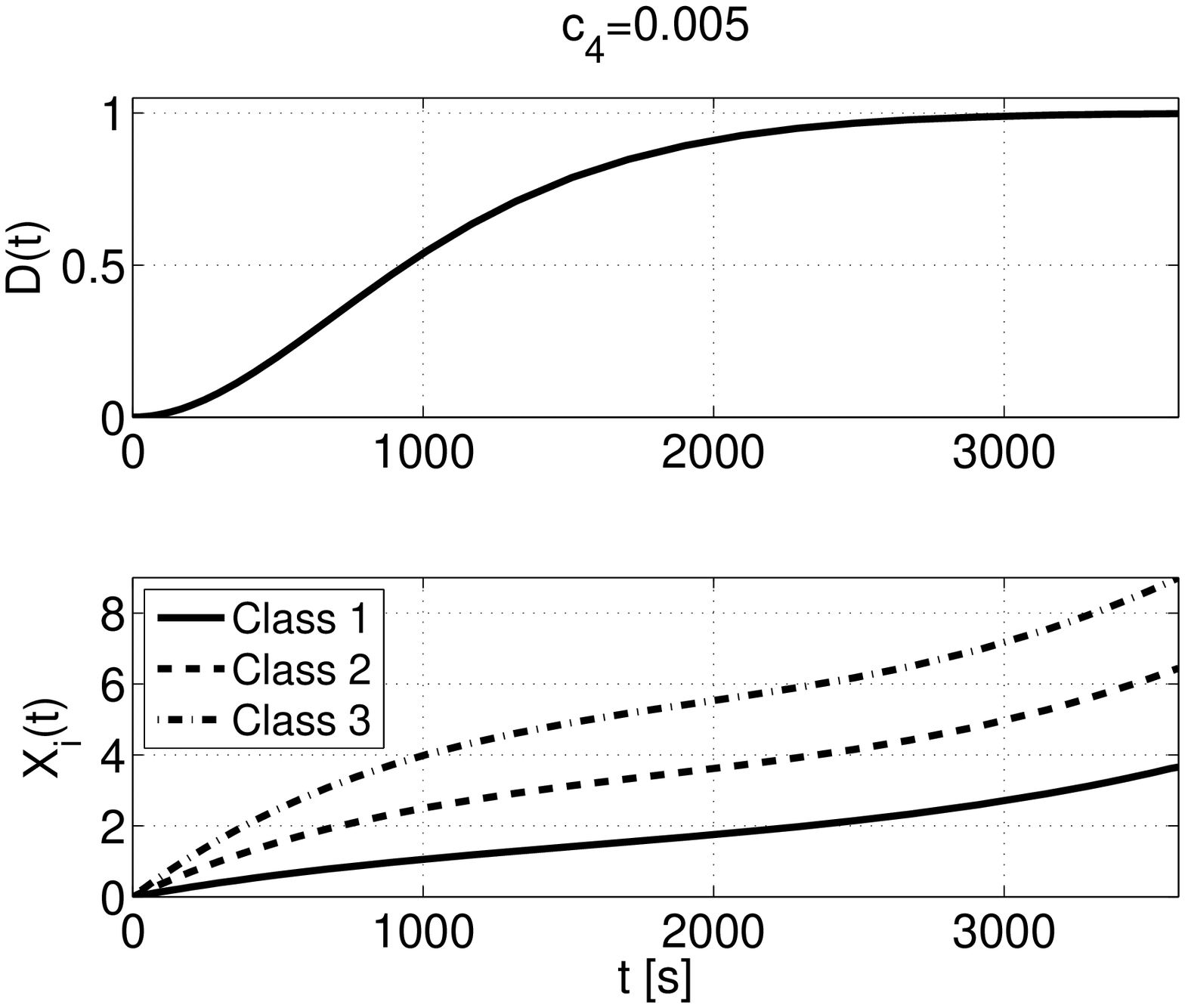}}
  \subfloat[]{\includegraphics [width=6.5cm]{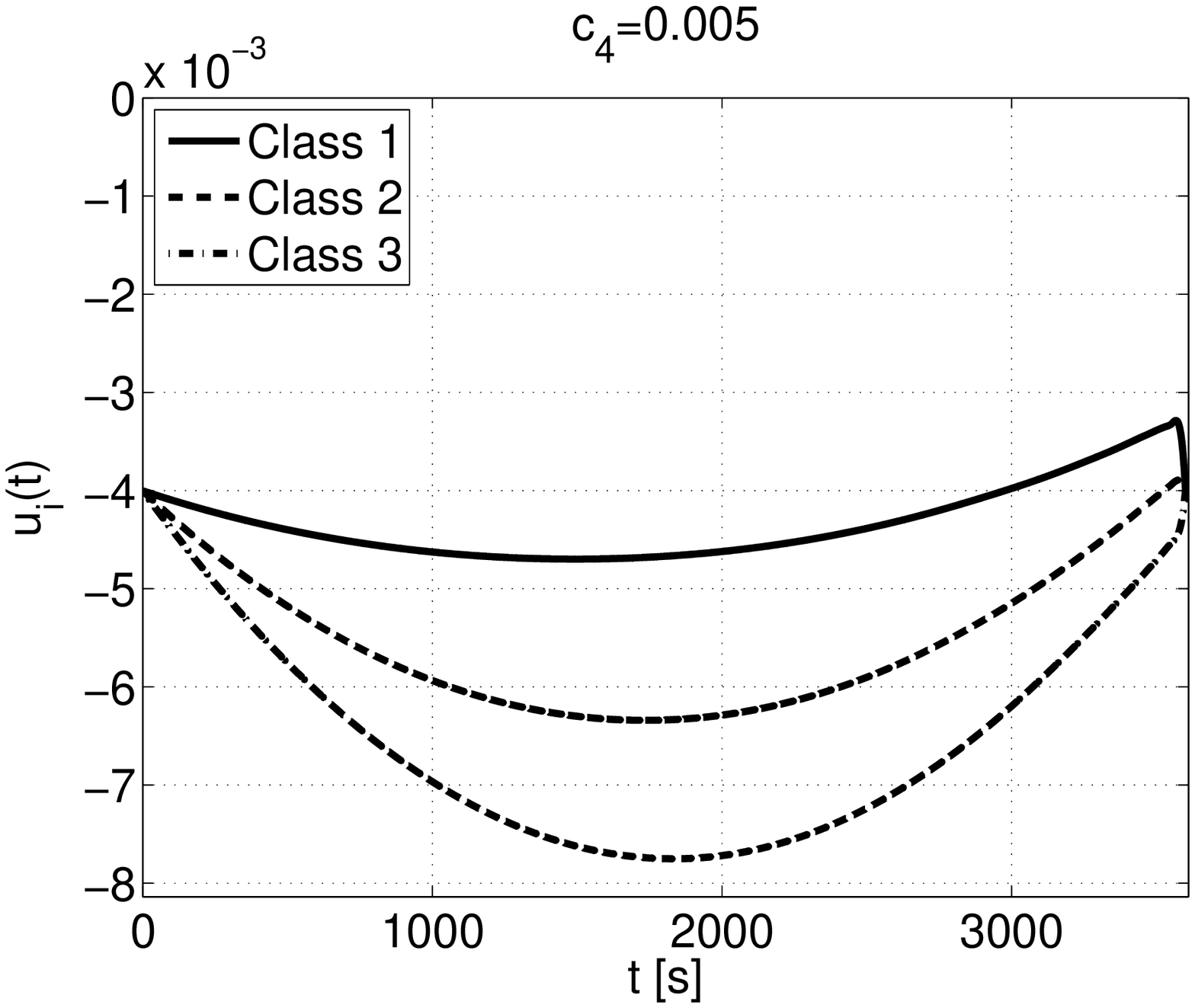}%
    \put(-100,100){\includegraphics [width=2.5cm]{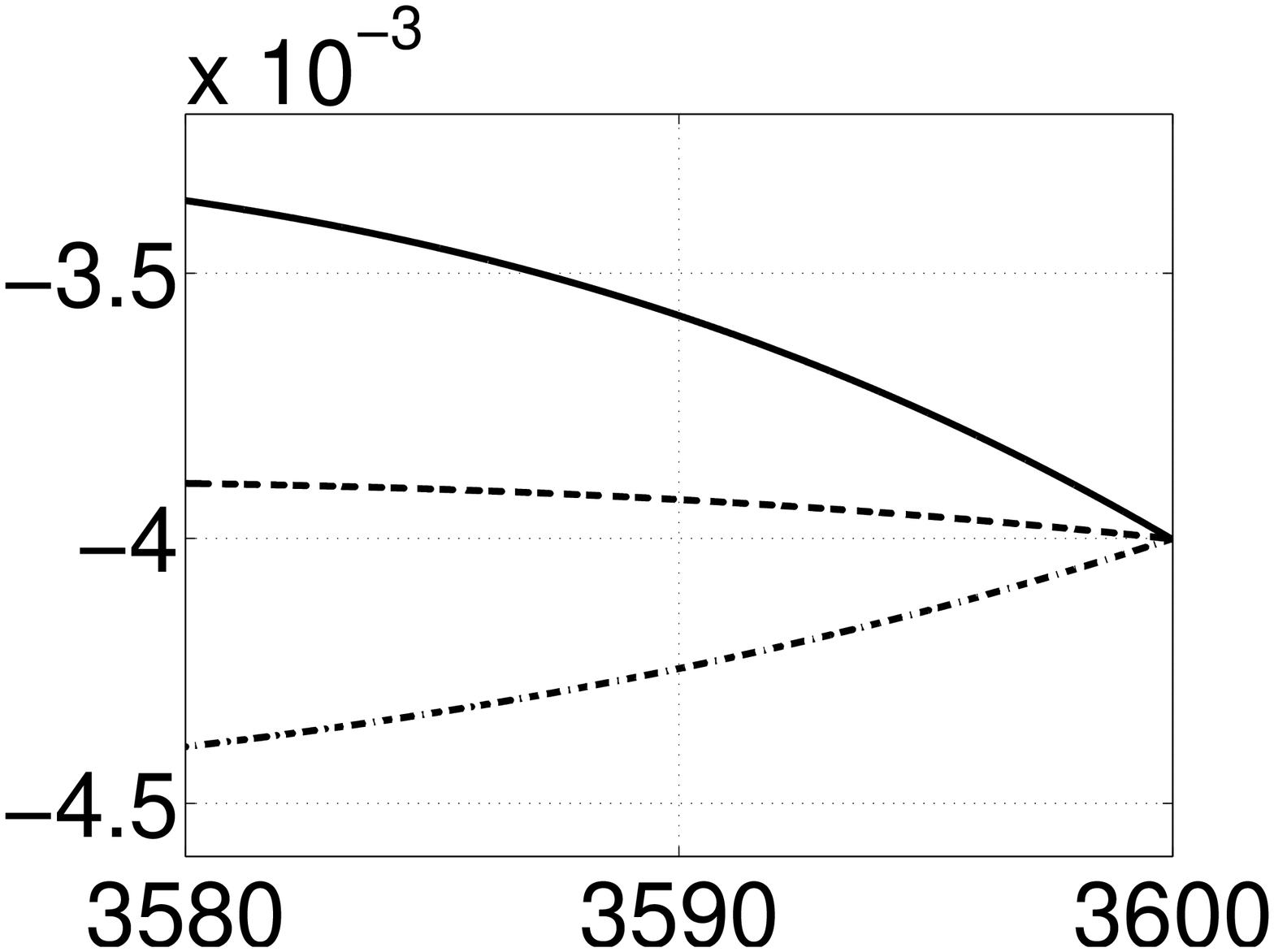}
               \put(-2,-2){\vector(3,-2){8.8mm}}}
}
\caption{Dynamical control of the timers, non-uniform case, $N=151$, $\tau=3600$ s; (a) delay CDF (upper) and dynamics
  of infected nodes per class $X_i(t)$ (lower); (b) control $u_i$. Parameters are as in Fig.~\ref{fig:nonuniform_O},
 $\lambda_{s1}=0.7 \lambda_{s2}$ and $\lambda_{s3}=1.3 \lambda_{s2}$; again $||\Lo||=||\Li||=\lambda_0$.%
}\label{fig:nonuniform_I}
\end{figure*}

In addition to the setting of Fig.~\ref{fig:nonuniform_O}, we considered also 
different intermeeting intensities with the source at different classes 
(see Fig~\ref{fig:nonuniform_I}). 
In addition to the setting described in the previous case, in particular, 
$\lambda_{s1}=0.7 \lambda_{s2}$ and $\lambda_{s3}=1.3 \lambda_{s2}$. As depicted there, the change of the
relative intermeeting rates within the three classes has a marked impact into the way the control is performed
compared to the case of Fig.~\ref{fig:nonuniform_O}.
In this case, in fact, the timeout rate, with respect to classes $2$ and $3$ is still larger, in order to limit the
increase of the number of messages; notice, though, that the infected nodes dynamics of those two classes is
basically the same of the previous case. But, the control of the timeouts for class $1$ is much milder than seen
previously, due to the lower intermeeting rate within that class, which reduces the need for high timeout
rates.

\subsection{The use of reference timeouts rates}

\begin{figure*}[t]
\centering
  \subfloat[]{\includegraphics [width=6.5cm]{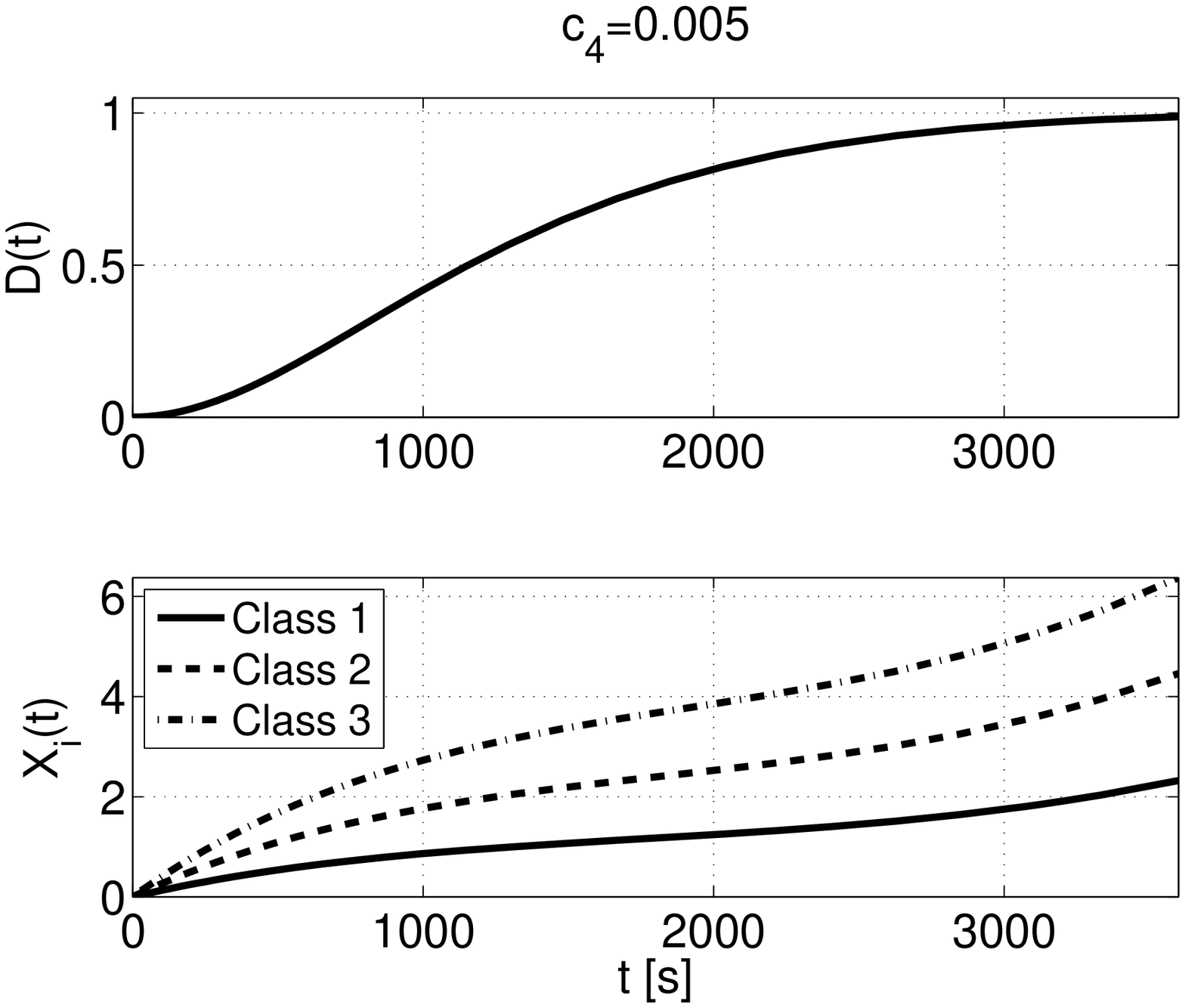}}
  \subfloat[]{\includegraphics [width=6.5cm]{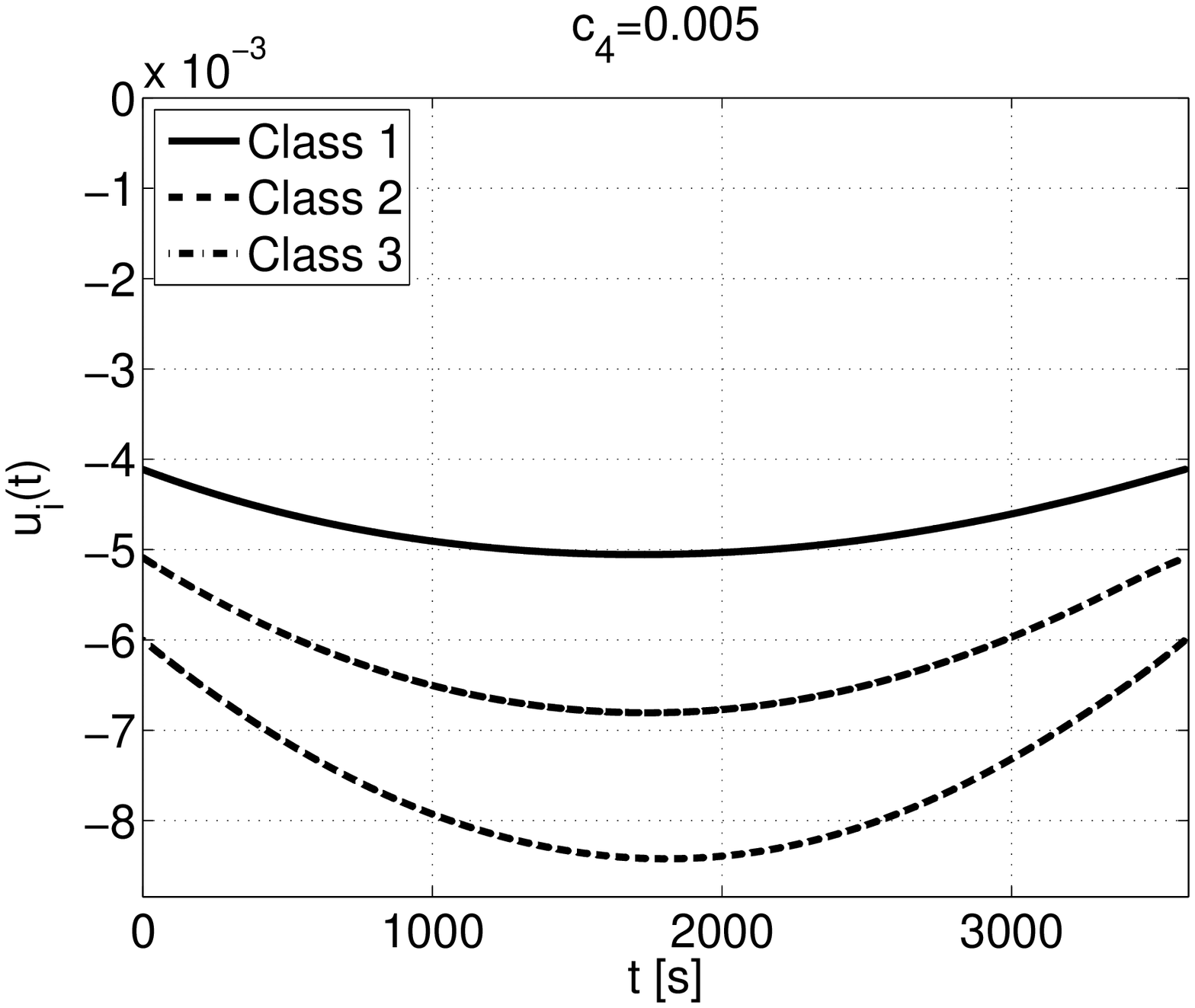}%
\put(-100,100){\includegraphics [width=2.5cm]{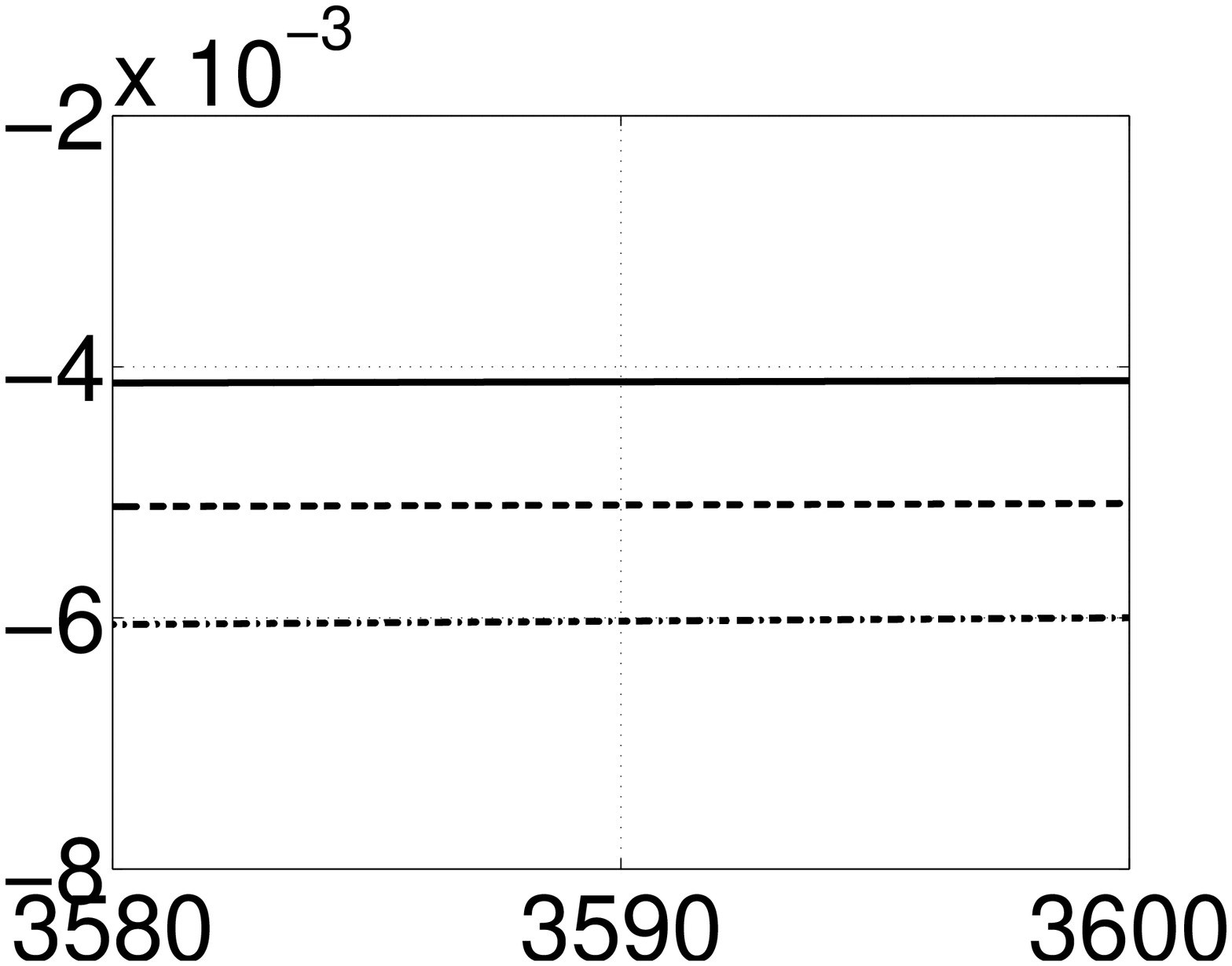}
\put(0,-2){\vector(1,-1){8.8mm}}}
}
\caption{Dynamical control of the timers, non-uniform case, $N=151$, $\tau=3600$ s; (a) delay CDF (upper) and dynamics
of infected nodes per class $X_i(t)$ (lower); (b) control $u_i$. Parameters are as in Fig.~\ref{fig:nonuniform_I},
$\overline u=-(4.1, 5.1, 6)^T\times 10^{-3}$.%
}\label{fig:nonuniform_u}
\end{figure*}

In the last two cases, i.e., Fig.~\ref{fig:nonuniform_O}b and Fig.~\ref{fig:nonuniform_I}b, we expanded the time scale
around $\tau$ in order 
to confirm the numerical stability of our solution. In particular, 
the final value of the control is dictated by the reference value $\overline u$. 
This also suggests that finer tuning of the control
can be obtained using different values for each $u_i$. With respect to Fig.~\ref{fig:nonuniform_u}b, we observe 
that this fine tuning is beneficial: under the same settings of Fig.~\ref{fig:nonuniform_I}b, the system is able 
to reach the desired high delivery probability,  whereas the number of copies in the system decreases compared to 
the use of uniform penalty on timeout rates.


\begin{remark}
The numerical evaluations represented before do not exhaust the range of the possible parameters of the
problem. For the choices showed before, we limited to cases where the optimal solution is compatible with the
constraints  $0\leq \sum X_i\leq N$ and $u_i\leq 0$; for cases of practical interest, due to the particular
structure of the two-hop routing protocol, the upper bound on the number of nodes is usually satisfied.
We found that a crucial parameter is the reference intermeeting intensity $\lambda_0$, which appears to strongly
impact the sensitivity of the optimal solution; in particular, it may determine settings where the constraint
$X_i\geq 0$ cannot be attained. More precisely, the approach seems to show some limitation when dealing with
small values of $\lambda_0$.
\end{remark}


\section{Conclusion}\label{sec:conclusion}


We have focused in this paper on controlling the spreading of
message under  two-hop relay forwarding. We exploited the linear form
of the dynamics of this regime to study the control problem within
the linear quadratic control framework. This allowed us to study
numerically the tradeoff achievable by tuning various parameters
that define the cost.\\
There exist several aspects that were not covered in the present work, which deserve 
further investigation. We showed that there exist constraints on the choice 
of optimization parameters; this relates to the need to keep the cost function well defined and bounded. 
We also remarked the need to satisfy the constraints on the dynamics of the number of infected nodes:
 to this respect, the numerical evaluations reported before do not exhaust the range of the 
possible parameters. Due to the particular structure of the two-hop routing protocol, the 
upper bound on the number of nodes is usually satisfied. We observed, though, that 
the reference intermeeting intensity $\lambda_0$ strongly affects the sensitivity of the optimal 
solution; in particular, it may determine settings where the constraint $X_i\geq 0$ cannot be 
attained, especially for very small values of $\lambda_0$. Notice, however, that 
even when the constraints are satisfied for the dynamics of the averages, 
they need not be satisfied each sample. An interesting question is to determine 
the probability that they are not satisfied by some realization.\\
From a practical standpoint, there are other directions for future work. The implementation of 
the control in two-hop routing can be done at the source only, and timers can be regulated 
through appropriate time-stamping of message copies (relays perform message discarding simply comparing the time 
elapsed since message generation and the control reported on the message copy). The crucial problem, 
conversely, is the estimation of the system parameters, namely, $N$ and $\lambda_{ij}$ at the source node. 
Also, previous work \cite{ANDM}, showed that in the $1$-dimensional case, when these parameters are unknown, 
it is still possible to obtain a policy that converges to the optimal one by using some auto-tuning mechanism. 
Using stochastic approximation this policy is shown to be optimal for the average cost criterion.

An interesting research direction would be to combine the optimal control techniques presented in this paper 
with coding techniques such as fountain codes or network coding \cite{code1,code2}.


\section{Acknowledgments}


This work has been partially supported by the European Commission within the framework of the BIONETS project IST-FET-SAC-FP6-027748, 
see \url{www.bionets.eu}. Research reported here has also been facilitated by a UIUC-INRIA Collaborative Research Grant jointly from
the University of Illinois at Urbana-Champaign and INRIA, France.




\section*{Appendix}\label{sec:appendix}


\subsection*{Deriving the success probabilities}

We follow the derivation in
\cite[Appendix A]{SH} which we extend so as to handle the multi-class case,
as well as to handle the case of non-homogeneous parameters (we allow
\[
Pr(T_d >t+h|T_d >t , {\cal F}_t )= 1 - h \sum_i \xi_i(t) \Lambda_{id} (t)+ o(h)
\]
which implies
\[
Pr(T_d >t+h | {\cal F}_t )= Pr(T_d >t | {\cal F}_t )( 1 - h \sum_i \xi_i(t)
\Lambda_{id} (t))
\]
As $\Psi (t) = 1 - Pr(T_d >t | {\cal F}_t )$ we get
\[
\frac{d \Psi (t) }{dt} = (1 - \Psi (t)) \sum_i \xi_i(t) \Lambda_{id} (t)
\]
Thus
\[
\frac{ \frac{d \Psi (t) }{dt} }{ 1 - \Psi (t) } =
\sum_i \xi_i(t) \Lambda_{id} (t)
\]
\[
- \log( 1 - \Psi (t)) =
\sum_{i=1}^K \Lambda_{i,d} \int_{s=0}^t \xi_i(s) ds + C_1
\]
and hence
\begin{equation}\label{delay}
\Psi (t) = 1 - C_2 \exp \Big( -
\sum_{i=1}^K \Lambda_{i,d} \int_{s=0}^t \xi_i(s) ds  \Big)
\end{equation}
For $t$ large, $\Psi (t)$ should tend to one, which it does. Also as $t\to 0$, $\Psi (t)$ should go to {\it zero}.
This last condition implies that $C_2=1$.


\subsection*{Proof of Prop.~\ref{prop:suffcond}}


\begin{proof}
$R>0$ if and only if $-c_1/c_3\Ld \Ld^T + I + c_4/c_3 \Lo^2>0$. 
Also, $||\Ld||^2$ is the only positive non-zero eigenvalue of $M = \Ld \Ld^t $, whose eigenvector
is $\Ld$. 
$M$ is symmetric, so let $V$ an orthogonal matrix such that
\[
\mbox{diag}(||\Ld||^2,0,\ldots,0)^T=V^T M V,
\]  
where in particular $V=(\Ld^T, c_2^T, \cdots, c_K^T)$ and $c_i\in \ker(M)$.

Also, $R>0$ iff  $\tilde R = V^TRV>0$, i.e.
\[
\tilde R=I-\frac{c_1}{c_3}\mbox{diag}(||\Ld||^2,0,\ldots,0)+\frac{c_4}{c_3} V^T \Lo\Lo V>0.
\]  
Let $e_1=(1,0,\ldots,0)$: the sufficient condition is obtained since if $e_1^T \tilde R e_1>0$, then $vRv>0$ for 
$\forall v \in \mathbb R^k$. Hence,  
\begin{eqnarray}\label{eq:suffcond}
e_1 \tilde R e_1^T&=& (1-\frac{c_1}{c_3})||\Ld||^2+\frac{c_4}{c_3} ||\Lo V e_1||^2 \nonumber \\
&=& (1-\frac{c_1}{c_3})||\Ld||^2+\frac{c_4}{c_3} ||\Lo \Ld||^2 \nonumber 
\end{eqnarray}
from which the statement follows.
\end{proof}


\subsection*{Calculation of $P$}


In the following we derive a closed form solution for $P$; we assume that $(\Lo)_{ii}>0$, 
for $i=1,2,\ldots,K$\footnote{The expression that we derive in the following requires $\Lo$ to 
be invertible but not diagonal.}. The solution of the matrix Riccati equation is given by 
$P=P_2P_1^{-1}$, where $P_1,P_2$ are $2K \times 2K$ matrices solutions of
\begin{equation}\label{lab:dynassociated}
{{\dot P_1} \choose {\dot P_2}} = H {P_1\choose P_2}, \qquad P_1(t_f)=\ik{2K}, \;P_2(t_f)=Q_f,
\end{equation}
where $\ik{2K}$ is the $2K \times 2K$ identity matrix; $P_1$ is guaranteed to be invertible 
for all $0\leq t\leq t_f$. In our case, if we choose $c_2=1$, the Hamiltonian matrix is
\begin{equation}\label{lab:hamiltonian}
H=\left( \begin{array}{cc} A & -BB^T \\ 0 & -A^T \end{array} \right )=\left( \begin{array}{cccc} -\Lo & 0 & -\ik{K} & 0 \\ \ik{K} & 0 & 0 & 0\\ 0 & 0  & \Lo & -\ik{K} \\ 0 & 0 & 0 & 0\end{array} \right )\nonumber%
\end{equation}

The associated dynamical system (\ref{lab:dynassociated}) solves for 
\[
{P_1(t) \choose P_2(t)} = e^{H(t-t_f)} { \ik{2K} \choose Q_f }
\]
so that we are interested in the explicit calculation of the matrix 
$e^{H(t-t_f)}= \sum_{k=0}^\infty \frac{(t-t_f)^k}{k!} H^k$ where for the sake of notation, $x=(t-t_f)$.

The $k$-th power of $H$ can be derived as follows
\begin{proposition}\label{eq:hk}
$H^0=\ik{4K}$, $H^1=H$, for $k>1$:
$$
H^k=\left( \begin{array}{cccc} -\Lo^{k} & 0 & -\Lo^{k-1} & 0 \\ \Lo^{k-1} & 0 & 0 & \Lo^{k-3}\\ 0 & 0  & \Lo^k & -\Lo^{k-1} \\ 0 & 0 & 0 & 0\end{array} \right ),\quad \mbox{if $k$ is odd} 
$$
$$
H^k=\left( \begin{array}{cccc} \Lo^{k} & 0 & 0 & \Lo^{k-2} \\ -\Lo^{k-1} & 0 & -\Lo^{k-2} & 0 \\ 0 & 0  & \Lo^k & -\Lo^{k-1} \\ 0 & 0 & 0 & 0\end{array} \right ),\quad \mbox{if $k$ is even} 
$$
\end{proposition}
The above formula can be easily verified by induction from 
(\ref{lab:hamiltonian}); hence, if $E_{ij}$ is the $ij$--th $K\times K$ 
block of $e^{Hx}$, $i,j=1,2,3,4$, we obtain
$$
E_{ij}=\sum_{k=0}^\infty \frac{x^k}{k!} (H^k)_{ij}
$$
where $(H^k)_{ij}$ the $K\times K$ $ij$--block of $H^k$, $i,j=1,2,3,4$.

From (\ref{eq:hk}), $E_{12}=E_{31}=E_{32}=E_{41}=E_{42}=E_{43}=0$; the diagonal entries are  
$E_{11}=E_{33}^{-1}=\sum_{k=0}^\infty \frac{(-\Lo x)^k}{k!}=e^{-\Lo x}$ whereas $E_{22}=E_{44}=\ik{K}$.
Non-zero off-diagonal entries require some calculations, which bring
\[
E_{21}=\Lo^{-1} \Big (\ik{K}-e^{-\Lo x}\Big ),\quad E_{34}=-\Lo^{-1} \Big (e^{\Lo x} - \ik{K}\Big ).
\] 
For ease of presentation, given diagonal matrix $\Lambda$, we define 
$\CH{\Lambda}$ and $\SH{\Lambda}$ the diagonal matrix such that $\CH{\Lambda}_{ii}=\cosh(\Lambda_{ii})$ and 
$\SH{\Lambda}_{ii}=\sinh(\Lambda_{ii})$, respectively. Hence, it follows
\[
E_{13}=-\Lo^{-1} \SH{\Lo x}\no,\quad  E_{24}=\Lo^{-3}(\SH{\Lo x} - x \Lo )\no
\]
\[
E_{14}=\Lo^{-2}(\CH{\Lo x} - \ik{K} )\no,\quad E_{23}=-\Lo^{-2}(\CH{\Lo x} - \ik{K} )\no
\]
Thus, we obtain 
\begin{small}
\begin{eqnarray}
&&\hskip-7mm P_2=\left( \begin{array}{cc} 0 & -\Lo^{-1} \Big (e^{\Lo x} - \ik{K}\Big )R \\ 0 & R 
\end{array} \right )\nonumber \\
&&\hskip-7mm P_1=\left( \!\!\begin{array}{cc} e^{-\Lo x} &  \Lo^{-2}(\CH{\Lo x} - \ik{K} )R \\ \Lo^{-1} \Big (\ik{K}-e^{-\Lo x}\Big )  & \ik{K} + \Lo^{-3}(\SH{\Lo x} - x \Lo )R
\end{array} \!\!\right )\nonumber
\end{eqnarray}
\end{small}

The inverse of $P_1$ can be obtained leveraging the block form and requiring $P_1P_1^{-1}=\ik{2K}$
$$
P_1\left( \begin{array}{cc} A_1 & A_2 \\ A_3  & A_4
\end{array} \right )=\left( \begin{array}{cc} \ik{K} & 0 \\ 0  & \ik{K}
\end{array} \right )
$$
from which we obtain 
$$
A_3=M^{-1}(x)\Lo^{-1}(\ik{K}-e^{\Lo x}), \quad A_4=M^{-1}(x)
$$
where $M_x=\ik{K}+\Lo^{-3}[(\SH{\Lo x} - x \Lo)$$ - (\CH{\Lo x} - \ik{K})(e^{\Lo x} - \ik{K})]R$.  

Finally, we obtain
\begin{eqnarray}\label{eq:finalriccati}
P_{11}&=&  \Lo^{-1}(\ik{K}-e^{\Lo x})RM_x^{-1}\Lo^{-1}(\ik{K}-e^{\Lo x})\nonumber \\
P_{12}&=&  \Lo^{-1}(\ik{K}-e^{\Lo x})M_x^{-1}R\nonumber\\
P_{21}&=& RM_x^{-1}\Lo^{-1}(\ik{K}-e^{\Lo x})\nonumber\\
P_{22}&=&R M_x^{-1} \nonumber
\end{eqnarray}

We notice that matrix $M_x$ is symmetric and invertible; in fact, let $f(x)=\sinh(x)-x+(\cosh(x)-1)(1-e^x)$; 
it follows that $\dot f=-(e^x-1)^2$, so that $[(\SH{\Lo x} - x \Lo) - (\CH{\Lo x} - \ik{K})$$(e^{\Lo x} - \ik{K})]>0$
since $-t_f<x<0$. Then, $v^TM(x)v>0$ for $\forall v\not =0$, so that $M(x)>0$. $P$ is symmetric, as expected.



\begin{thebibliography}{99}


\bibitem{ANA}
A. Al-Hanbali, P. Nain, and E. Altman,
``Performance of ad hoc networks with two-hop relay routing and
limited packet lifetime'', \emph{Proc. of Valuetools}, Pisa, Italy, October 11-13, 2006.

\bibitem{ABD}
E.~Altman, T.~{Ba\c sar}, and F.~{{De} Pellegrini}, ``Optimal monotone forwarding 
policies in delay tolerant mobile ad-hoc networks,'' \emph{Elsevier Performance Evaluation}, 
article in press, \url{doi:10.1016/j.peva.2009.09.001}.

\bibitem{bcfh}
R. Bakhshi, L. Cloth, W. Fokkink and B.  R. Haverkort,
``Mean-Field analysis for the evaluation of gossip protocols'',
 \emph{ACM SIGMETRICS Performance Evaluation Review}, Vol. 36 , no. 3,  Dec 2008, pp. 32--39.

\bibitem{BMDP}
A. Bemporad, M. Morari, V. Dua and E. N. Pistikopoulos,
``The explicit linear quadratic regulator for constrained systems,'' 
{\it Automatica} 38, 3-20, 2002.

\bibitem{Levine_infocom2006}
J.~Burgess, B.~Gallagher, D.~Jensen, and B.~N. Levine, ``Maxprop: Routing for
  vehicle-based disruption-tolerant networking,'' \emph{Proc. of INFOCOM}, Barcelona, Spain, April 23--29, 2006.

\bibitem{Diot_infocom2006}
A.~Chaintreau, P.~Hui, J.~Crowcroft, C.~Diot, R.~Gass, and J.~Scott, ``Impact
  of human mobility on the design of opportunistic forwarding algorithms,'' 
\emph{Proc. of INFOCOM}, Barcelona, Spain, April 23--29, 2006.


\bibitem{fall_implementingDTN}
M.~Demmer, E.~Brewer, K.~Fall, S.~Jain, M.~Ho, and R.~Patra, ``Implementing
  delay tolerant networking,'' Intel, Tech. Rep. IRB-TR-04-020, 28 Dec. 2004.

\bibitem{FBS}
A.~E. Fawal, J.-Y.~L. Boudec, and K.~Salamatian, ``Performance analysis of self
  limiting epidemic forwarding,'' EPFL, Tech. Rep. LCA-REPORT-2006-127, 2006.

\bibitem{GNK}
R. Groenevelt, P.  Nain, and G.  Koole, ``The message delay in mobile ad
hoc networks'', \emph{in posters ACM SIGMETRICS 2005, Canada, 2005.}
\bibliographystyle{IEEE}

\bibitem{tse_mobility02}
M.~Grossglauser and D.~Tse, ``Mobility increases the capacity of ad hoc
  wireless networks,'' \emph{IEEE/ACM Trans. on Networking}, vol.~10, no.~4, 
 Aug. 2002, pp. 477--486.

\bibitem{gupta_capacity}
P.~Gupta and P.~R. Kumar, ``The capacity of wireless networks,'' \emph{IEEE
  Trans. on Information Theory}, vol.~46, no.~2, pp. 388--404, March 2000.

\bibitem{fall_routDTNs}
S.~Jain, K.~Fall, and R.~Patra, ``Routing in a delay tolerant network,''
  \emph{SIGCOMM Comp. Comm. Rev.}, vol.~34, no.~4, pp. 145--158, Oct. 2004.

\bibitem{GL}
G.~Leitmann, \emph{Optimal Control}, McGraw-Hill, 1966.

\bibitem{BO}
T. Ba\c sar and G. J. Olsder, \emph{Dynamic Noncooperative Game Theory}, SIAM Series in Classics in Applied 
Mathematics, SIAM, Philadelphia PA, USA, 1999.

\bibitem{MM}
M.~Musolesi and C.~Mascolo, ``Controlled Epidemic-style Dissemination Middleware for Mobile Ad Hoc Networks,'' 
\emph{Proc. of ACM Mobiquitous}, San Jose, California July  17-21, 2006.

\bibitem{SH}
T. Small and Z. J. Haas, ``The shared wireless infostation model  - a new ad hoc networking paradigm,''
{\it Proc. of  ACM MobiHoc}, Annapolis, Maryland, USA, June 1-3, 2003,

\bibitem{Zegura_Mobihoc06}
M.~M.~B. Tariq, M.~Ammar, and E.~Zegura, ``Message ferry route design for
  sparse ad hoc networks with mobile nodes,'' \emph{Proc. of ACM MobiHoc},
  Florence, Italy, May 22--25, 2006, pp. 37--48.

\bibitem{VB}
A.~Vahdat and D.~Becker, ``Epidemic routing for partially connected ad hoc
  networks,'' Duke University, Tech. Rep. CS-2000-06, 2000.

\bibitem{ZNKT}
X.~Zhang, G.~Neglia, J.~Kurose, and D.~Towsley, ``Performance modeling of
  epidemic routing,''  \emph{Elsevier Computer Networks}, vol. 51, no. 10, July 2007, pp. 2867-2891.

\bibitem{ANDM}
E.~Altman, G.~Neglia, F.~De Pellegrini, and D.~Miorandi, ``Decentralized
  stochastic control of delay tolerant networks,'' \emph{Proc. of IEEE INFOCOM},
  Rio de Janeiro, April 15-19, 2009.

\bibitem{code1}
E.~Altman and  F.~De Pellegrini, ``Forward Correction and Fountain codes in Delay Tolerant Networks,'' 
\emph{Proc. of IEEE INFOCOM}, Rio de Janeiro, April 15-19, 2009.

\bibitem{code2}
E. Altman, F. De Pellegrini, and L. Sassatelli, ``Dynamic control of Coding in Delay Tolerant Networks,'' 
available on arXiv
\end{thebibliography}
\end{document}